\RequirePackage{etoolbox}
\csdef{input@path}{%
 {sty/}
 {img/}
}%
\csgdef{bibdir}{bib/}

\documentclass{article}

\newcommand{\E}{\mbox{E}}
\newcommand{\inb}{\mbox{INB}}
\newcommand{\V}{\mbox{Var}}

\usepackage{bm}
\usepackage{url}
\usepackage{graphicx}
\usepackage{lipsum}
\usepackage{verbatim}
\usepackage{multicol}
\usepackage{xcolor}
\usepackage{multirow}
\usepackage{tikz}
\usetikzlibrary{positioning}

\author{Anna Heath, Ioanna Manaolopoulou and Gianluca Baio}

\begin{document}


\title{Estimating the Expected Value of Sample Information across Different Sample Sizes using Moment Matching and Non-Linear Regression}

\maketitle

\begin{abstract}
Background: The Expected Value of Sample Information (EVSI) determines the economic value of any future study with a specific design aimed at reducing uncertainty in a health economic model. This has potential as a tool for trial design; the cost and value of different designs could be compared to find the trial with the greatest net benefit. However, despite recent developments, EVSI analysis can be slow especially when optimising over a large number of different designs. Methods: This paper develops a method to reduce the computation time required to calculate the EVSI across different sample sizes. Our method extends the moment matching approach to EVSI estimation to optimise over different sample sizes for the underlying trial with a similar computational cost to a single EVSI estimate. This extension calculates posterior variances across the alternative sample sizes and then uses Bayesian non-linear regression to calculate the EVSI. Results: A health economic model developed to assess the cost-effectiveness of interventions for chronic pain demonstrates that this EVSI calculation method is fast and accurate for realistic models. This example also highlights how different trial designs can be compared using the EVSI. Conclusion: The proposed estimation method is fast and accurate when calculating the EVSI across different sample sizes. This will allow researchers to realise the potential of using the EVSI to determine an economically optimal trial design for reducing uncertainty in health economic models. Limitations: Our method relies on some additional simulation, which can be expensive in models with very large computational cost.
\end{abstract}

\section{Introduction}
The Expected Value of Sample Information (EVSI) \cite{RaiffaSchlaifer:1961} quantifies the expected economic benefit of a data collection exercise with a given design. In general, new data update information about the parameters underlying a health economic model. This, in turn, updates information about the optimal treatment or intervention and thereby reduces the possibility of funding an inefficient treatment, which may only have been deemed to be economically efficient due to deficiencies in the current information.

The EVSI is a Value of Information (VoI) measure \cite{Howard:1966} used to analyse the potential benefit of resolving uncertainty in a decision model. Amongst these measures, the EVSI has the greatest potential to inform research decisions as it quantifies the economic benefit of a specific study. This value can be compared directly with the cost of the trial to determine whether a study should go ahead, making the EVSI an important tool for decision makers and funders looking to efficiently allocate research and development resources \cite{McKennaClaxton:2011}. 

In addition to this, the EVSI has potential as a tool for trial design when it is calculated for a number of different designs \cite{WillanPinto:2005}. The trial with the largest ``net'' economic value is then selected, by considering the cost of implementation alongside the EVSI. Using the EVSI in this setting ensures that the data collected in the study are sufficient to reliably inform the health economic model. Therefore, following the study, the cost-effectiveness of the different treatments can be correctly assessed, which is a key requirement within publicly funded health systems where treatments/interventions must demonstrate cost-effectiveness over competitors \cite{Claxtonetal:2005,eunethta:2014,CADTH:2006,Australia:2008}. However, the computational time required to calculate the EVSI based on realistic health economic models has restricted its practical applications \cite{Steutenetal:2013} meaning that studies are normally designed using power calculations to determine an appropriate sample size.

With the exception of cases in which the health economic model conforms to certain conditions \cite{Adesetal:2004,Weltonetal:2014,BrennanKharroubi:2007b}, traditionally, the EVSI has been estimated using computationally intensive nested Monte Carlo simulations \cite{Migueletal:2016,AndronisBarton:2016,Pandoretal:2015,Weltonetal:2011}. However, recent methodological advances have employed various methods to avoid the need for full nested simulations, irrespective of the underlying model structure \cite{Strongetal:2015,Menzies:2016,Jalaletal:2015,Heathetal:2017b,JalalAlarid:2018}. In theory, these methods have unlocked the power of EVSI analysis for use in research prioritisation and study~design.

Unfortunately, using the EVSI as a tool to identify the optimal study design still requires calculations to be made across a number of different possible data collection exercises, e.g.~changing the sample size or the main outcome. In this paper, we present an extension to the ``moment matching'' estimation method \cite{Heathetal:2017b,Heathetal:2018} that calculates the EVSI across different sample sizes for the future study with approximately the same computational cost as a single estimate.

Our extension uses Bayesian non-linear regression to estimate the EVSI across different sample sizes, by incorporating prior information about the behaviour of EVSI. It also models uncertainty in the EVSI that arises from its estimation using simulation. The resulting quantification of uncertainty in the EVSI estimate allows funders to assess the accuracy of their research and development decisions. 

The moment matching method only requires that the health economic model is based on a Bayesian model that incorporates the data arising from the study. This requirement comes directly from the definition of the EVSI and imposes no restrictions on the complexity of the health economic model or data collection exercise used to calculate the EVSI. Additionally, this requirement allows EVSI calculations to be undertaken whilst considering common trial issues such as missingness or loss to drop out in the data. Therefore, this extension to the moment matching method can find the optimal sample size while taking into account realistic assumptions about the study and the model.

To present this extension, we briefly introduce the EVSI in \S\ref{EVSI - section} and the standard moment matching method \cite{Heathetal:2017} in \S\ref{Method - Heath}. Following this, we present a Bayesian non-linear model that can be used to estimate the EVSI across different sample sizes based on one simulation per sample size in \S\ref{sample-sizes}. Finally, we then present a practical implementation of this method using a real-life health economic model developed to compare treatments for chronic pain \cite{Sullivanetal:2016}. We compare our estimates with the computationally intensive nested Monte Carlo simulation method to demonstrate its accuracy. We also explore how the EVSI can be used as a tool for trial design by considering two alternative designs across a range of sample sizes. 

\section{The Expected Value of Sample Information}\label{EVSI - section}

The EVSI quantifies the economic benefit of reducing uncertainty in a health economic model using a specific method for data collection. Therefore, we first define a health economic model to compare treatments for a specific disease/health state. This is typically defined in terms of a large vector of parameters $\bm\theta$, used to describe the properties of the underlying disease, for example survival probabilities, Quality of Life measures and health service costs. In a Bayesian setting, a joint probability distribution $p(\bm\theta)$ defines the current uncertainty level in these parameters and is typically informed by a number of different data sources such as clinical trials augmented by additional information from the literature in a process known as Probabilistic Sensitivity Analysis (PSA)~\cite{BaioDawid:2011}. The corresponding frequentist bootstrap approach to PSA typically models each parameter in $\bm\theta$ individually to represent the current level of uncertainty in the model parameters.

In a health economic model, the value of each treatment is normally defined using a ``net benefit'' function dependent on the model parameters, NB$^{\bm\theta}_t$ for $t=1,\dots,T$ possible interventions/alternatives. The probability distribution $p(\bm\theta)$ then induces a distribution for the net benefit for each treatment. Typically, we consider that variation due to individual level response to treatments had been marginalised out, so if the parameters were known exactly then the net benefit of each treatment would also be known. The optimal treatment given current evidence is then found by averaging out all uncertainty for each treatment, $\mathcal{NB}_t=\mbox{E}_{\bm\theta}\left[\mbox{NB}^{\bm\theta}_t\right]$, and then finding the most valuable treatment, $\max_t \mathcal{NB}_t$.

However, the current uncertainty in the parameters may suggest that the incorrect treatment is optimal. This is because additional information gleaned from a study would update the distribution of the model parameters which would then update the distribution of the net benefit. This could change the optimal treatment, indicating that the current information is not sufficient to make the correct treatment recommendation, for example, additional side effects could be observed when a new treatment is monitored for a longer period of time in a greater number of patients. In this case, the additional data have value as patients are prevented from receiving an inefficient treatment and financial resources are saved.

Formally, a study collects data, denoted $\bm X$, which update the information about the model parameters $\bm\theta$. This updating falls within the Bayesian paradigm in which we consider $p(\bm\theta)$ as the prior distribution for the model parameters. The sampling distribution for the data is combined with the prior to determine a \emph{posterior} for model parameters $p(\bm\theta\mid\bm X)$. The net benefit for each treatment is calculated as a function of $\bm\theta$ and the optimal treatment found by maximising the posterior expected net benefit. The economic value of the study is then the difference between the expected net benefit of the optimal treatment with the updated information and the current information. 

However, as the EVSI is calculated \emph{before} the data have been collected, the study outcome is unknown. Therefore, we consider a \emph{distribution} for all the possible datasets that could arise from the study and calculate the value of every study outcome. The EVSI is the average study value over all the possible future datasets. Mathematically, this is expressed as
\begin{equation}\mbox{EVSI} = \E_{\bm X}\left[\max_t\left\{\E_{\bm\theta\mid\bm X}\left[\mbox{NB}^{\bm\theta}_t\right]\right\}\right] - \max_t\left\{\E_{\bm\theta}\left[\mbox{NB}^{\bm\theta}_t\right]\right\},\label{EVSI}\end{equation} 
where the distribution of $\bm X$ should be consistent with how the data would be analysed if the study were run. This means that we define a sampling distribution for the data given the parameters, $p(\bm X\mid\bm\theta)$, which is combined with $p(\bm\theta)$ to give the distribution of all the possible samples.

It is almost impossible to calculate the EVSI analytically, so in practical scenarios it is estimated using simulations. Computationally, the challenge is to compute the inner ``posterior'' expectation for the net benefit of each treatment for each possible study, $\E_{\bm\theta\mid\bm X}\left[\mbox{NB}^{\bm\theta}_t\right]$ for $t=1,\dots,T$. If this posterior expectation is available in closed form, then estimating the EVSI by simulation requires little computational effort as we simply require a large number of simulated study outcomes. However, in practical models this analytic expectation will rarely be available so alternative methods must be used. The most general method is to estimate the inner expectation using simulations from the posterior of $\bm\theta$, conditional on each of the simulated study outcomes. This means that the simulations required to estimate the posterior expectation are \emph{nested} within the simulations of possible study outcomes, leading to a nested Monte Carlo simulation~method with relatively high computational cost as typically around 1000 simulated study outcomes are used. 

Our ``moment matching'' method also estimates the EVSI using simulation but circumvents the need for full nested Monte Carlo simulation by estimating the ``posterior'' expectation across all the potential samples using a small number, between 30-50, of simulated future datasets \cite{Heathetal:2017b}.

\section{The moment matching method}\label{Method - Heath}
Thus far, we have presented the EVSI for health economic models that compare $T$ treatments. However, for simplicity, we restrict ourselves to the dual-decision setting, i.e.~$t=1,2$, before returning to the general setting in the supplementary material. To simplify the calculations still further, we work directly with the incremental net benefit, i.e.~$\mbox{INB}^{\bm\theta}=\mbox{NB}^{\bm\theta}_1-\mbox{NB}^{\bm\theta}_2$. The incremental net benefit contains all the information required to make a decision between treatment 1 and treatment 2. Specifically, treatment 1 is optimal if the expected incremental net benefit is positive and treatment 2 is optimal otherwise. Therefore, to calculate the EVSI, we must estimate the posterior expectation of INB$^{\bm\theta}$ across the different study outcomes, which we denote $\mu^{\bm X}=\E_{\bm\theta\mid\bm X}\left[\mbox{INB}^{\bm\theta}\right]$. To further clarify the moment matching method, a full list of notation is given in Table~\ref{notation} at the end of \S\ref{sample-sizes}.

\subsection{Estimating the distribution of $\mu^{\bm X}$ using moment matching}
Formally, the moment matching method is used to estimate the \emph{distribution} of $\mu^{\bm X}$, denoted $p(\mu^{\bm X})$, which is induced by the distribution of the potential datasets $\bm X$. Specifically, the moment matching method exploits the similarity between $p\left(\mu^{\bm X}\right)$ and the distribution of a specific function of INB$^{\bm\theta}$ to reduce the required number of nested Monte Carlo simulations \cite{Heathetal:2017b}. To find this function of INB$^{\bm\theta}$, we assume that the future sample only directly updates a subset $\bm\phi$ of the model parameters $\bm \theta$; for example, a clinical trial would inform the primary clinical outcome but is unlikely to give information on the societal costs of the disease. The distribution of $\mu^{\bm X}$ is then expected to be similar to the distribution of \begin{equation}\mbox{INB}^{\bm\phi} = \E_{\bm\theta\mid\bm\phi}\left[\mbox{INB}^{\bm\theta}\right], \label{EVPPI}\end{equation} except that the \emph{variance} of $\mu^{\bm X}$ is smaller than the variance of INB$^{\bm\phi}$ \cite{Heathetal:2017b}. In practice, therefore, the moment matching method approximates $p\left(\mu^{\bm X}\right)$ by linearly rescaling the simulated values of $\inb^{\bm\phi}$ to reduce their variance so it is equal to the variance of $\mu^{\bm X}$, denoted $\sigma^2_{\bm X}$, which must also be estimated. 

\subsection{Estimating simulated values of $\inb^{\bm\phi}$}
Simulated values of $\mbox{INB}^{\bm\phi}$, denoted INB$^{\bm\phi_s}$ for $s=1,\dots,S$, can be found in a computationally efficient manner using methods developed for the calculation of the Expected Value of Partial Perfect Information (EVPPI), which is based on the expectation in equation (\ref{EVPPI}). The main reason for using these methods is that the EVPPI is an upper bound for the EVSI and is much simpler to compute. As the EVSI must be \emph{large} to indicate that a study is worthwhile, a study with a small EVPPI, the upper limit of the value, will not be economically viable, irrespective of design. Therefore, the EVPPI should always be calculated before the EVSI as a screening procedure \cite{Tuffahaetal:2016} implying that INB$^{\bm\phi_s}$ should be available before attempting to calculate the EVSI.

In most health economic models, an EVPPI calculation method based on non-parametric regression is fast and accurate \cite{Heathetal:2017}. This uses regression to estimate INB$^{\bm\phi}$ and is easily generalisable implying that general purpose software \cite{BCEA:2017} or stand alone functions \cite{Strong:2012:Code} can estimate $\inb^{\bm\phi}$ based solely on the PSA simulations for $\bm\phi$ and $\inb^{\bm\theta}$.

\subsection{Estimating the variance of $\mu^{\bm X}$}
The variance of  $\mu^{\bm X}$, denoted $\sigma^2_{\bm X}$, is estimated using the law of total variance\[\sigma^2_{\bm X} = \V_{\bm X}\left[\E_{\bm\theta\mid\bm X}\left[\inb^{\bm\theta}\right]\right]=\V\left[\mbox{INB}^{\bm\theta}\right] -\E_{\bm X}\left[\V_{\bm\theta\mid\bm X}\left[\mbox{INB}^{\bm\theta}\right]\right].\] Firstly, the variance of $\inb^{\bm\theta}$, denoted $\sigma^2$, can be estimated by the sample variance of the PSA simulations of $\inb^{\bm\theta}$. Therefore, $\sigma^2_{\bm X}$ can be estimated by determining the expected posterior variance of the incremental net benefit, where the expectation is taken across the different potential datasets.

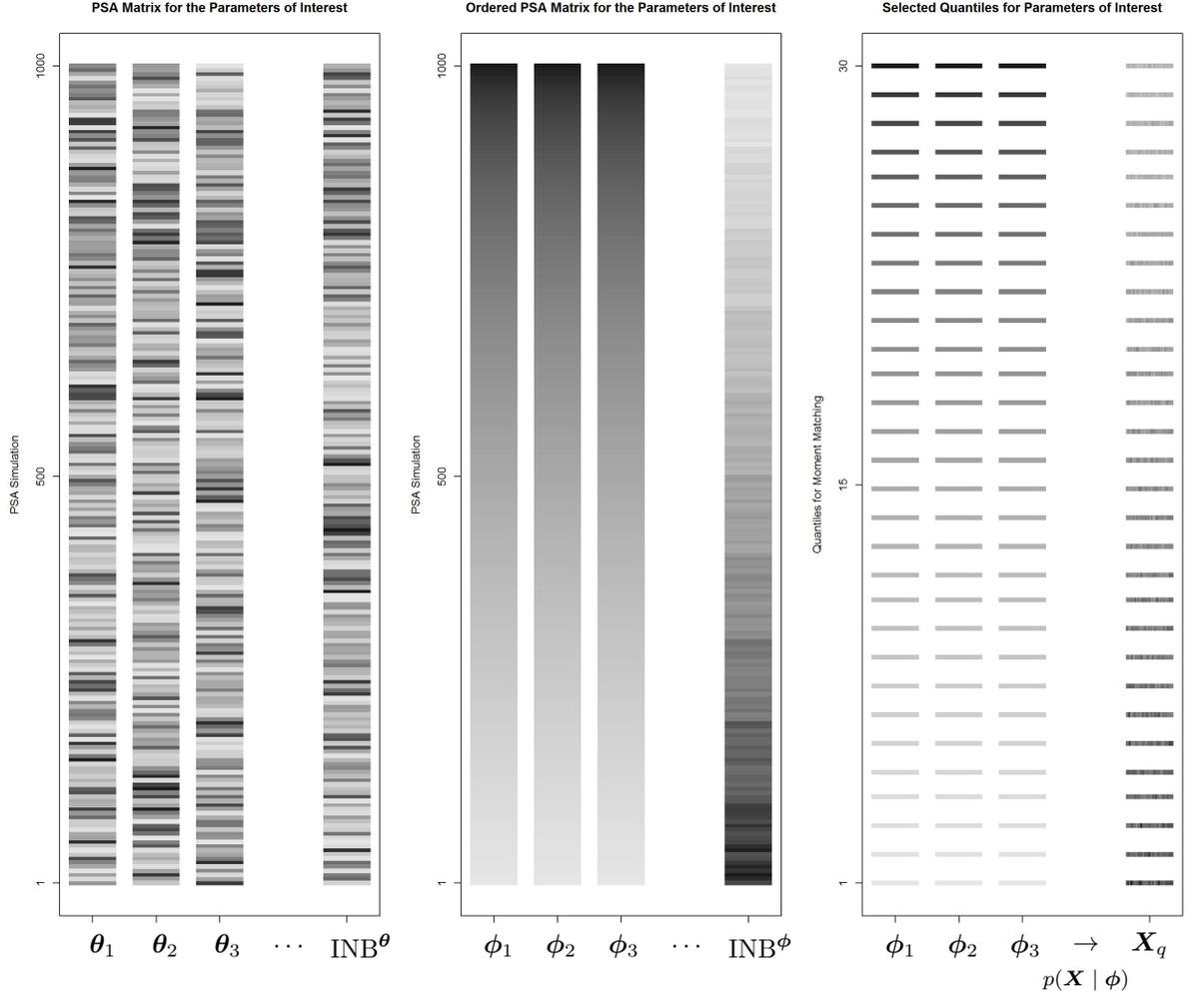
\begin{figure}[!h]
\begin{tikzpicture}
\node at (0,0) (a) {\includegraphics[width=16cm]{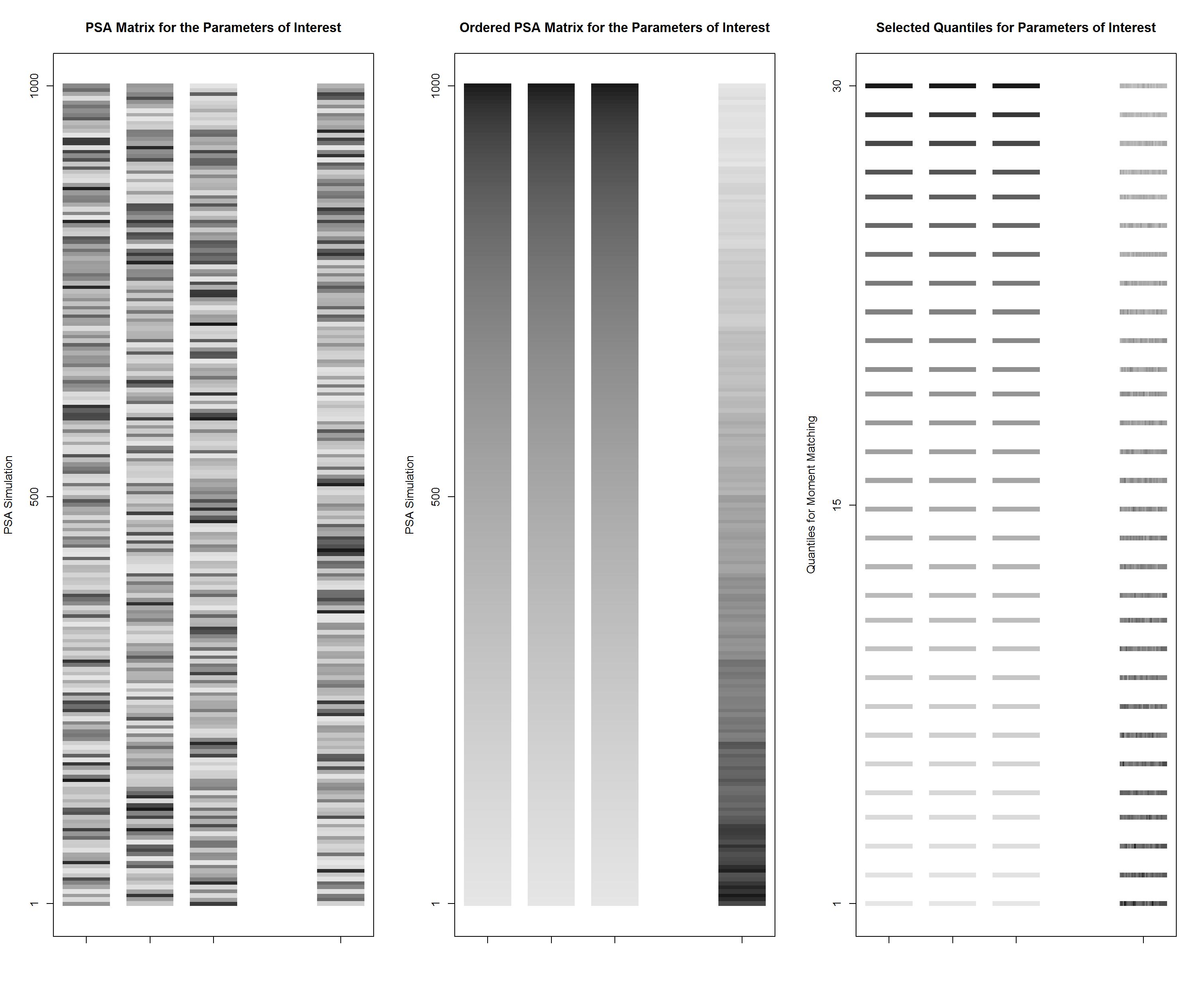}};
\node[below left =-2.5em and -4.85em of a,color=black] (b) {$\bm\theta_1$};
\node[right =.65em of b,color=black] (c) {$\bm\theta_2$};
\node[right =.65em of c,color=black] (d) {$\bm\theta_3$};
\node[right =.5em of d,color=black] (e) {$\dots$};
\node[right =.3em of e,color=black] (f) {$\inb^{\bm\theta}$};

\node[right = 2.75em of f,color=black] (g) {$\bm\phi_1$};
\node[right =.5em of g,color=black] (h) {$\bm\phi_2$};
\node[right =.5em of h,color=black] (i) {$\bm\phi_3$};
\node[right =.5em of i,color=black] (j) {$\dots$};
\node[right =.3em of j,color=black] (k) {$\inb^{\bm\phi}$};

\node[right = 2.8em of k,color=black] (l) {$\bm\phi_1$};
\node[right =.5em of l,color=black] (m) {$\bm\phi_2$};
\node[right =.5em of m,color=black] (n) {$\bm\phi_3$};
\node[right =.5em of n,color=black] (o) {$\rightarrow$};
\node[right =.5em of o,color=black] (p) {$\bm X_q$};
\node[below = 0em of o,color=black] (q) {\footnotesize$p(\bm X\mid\bm\phi)$};
\end{tikzpicture}
\caption{A pictorial representation of the simulation method used to estimate the posterior variances $\sigma^2_q$ for $q=1,\dots,Q$. The left-hand panel represents the standard PSA procedure, darker lines represent large values for the variable of interest while lighter lines represent smaller values. The middle panel selects the parameters of interest and reorders them smallest to largest. The right-hand panel demonstrates that the quantiles for $\bm\phi$ should be selected and then a dataset should be simulated for each row.}
\label{MMdescription}
\end{figure}

This expected posterior variance for INB$^{\bm\theta}$ can actually be estimated accurately using a small number $Q$ of nested simulations, with $30<Q<50$, provided the nested samples are undertaken using a specific method, demonstrated pictorially in Figure \ref{MMdescription}. This method begins with a standard PSA procedure, represented in the left-hand panel in Figure \ref{MMdescription} with $S=1000$. In this procedure, the parameters $\bm\theta$ are simulated from $p(\bm\theta)$ and then used to calculate the INB$^{\bm\theta}$. 

To proceed, we select the columns of the PSA matrix that contain the simulations for the parameters of interest $\bm\phi$. These columns are then ordered separately, represented by the middle panel in Figure \ref{MMdescription}. Finally, $Q$ equally spaced values are selected from these ordered lists to find the sample quantiles for each element of $\bm\phi$. For each row of sample quantiles, $\bm\phi_q$, $q=1,\dots,Q$, \emph{one} sample is simulated from $p(\bm X\mid \bm\phi_q)$, indicated by  $\bm X_q$, represented in the right hand graphic. To estimate the expected posterior variance, each of these $Q$ simulated future samples, $\bm X_q$, is used to update the posterior for the model parameters $p(\bm\theta\mid\bm X_q)$ and thereby induces a posterior distribution for INB$^{\bm\theta}$. In turn, this is estimated using simulations, with which the posterior variance of INB$^{\bm\theta}$, denoted  $\sigma^2_q$, can be estimated for each future sample. Finally, the variance $\sigma^2_{\bm X}$ is estimated by taking the average of the $\sigma^2_q$ and subtracting this from $\sigma^2$;  \[\sigma^2_{\bm X} =  \sigma^2 - \frac{1}{Q}\sum_{q=1}^Q \sigma_q^2.\] 

\subsection{Linear rescaling of INB$^{\bm\phi}$ to estimate the EVSI}\label{EVSI}
Once $\sigma_{\bm X}^2$ has been estimated, the EVSI is calculated by rescaling $\inb^{\bm\phi_s}$ so the variance of these simulated values is equal to $\sigma_{\bm X}^2$. This is achieved by first calculating \begin{equation}\eta^{\bm X}_s =\frac{\inb^{\bm\phi_s} - \mu}{\sqrt{\sigma^2_{\bm\phi}}} \sqrt{\sigma^2_{\bm X}} + \mu,\label{rescale}\end{equation} where $\mu=\E\left[\inb^{\bm\theta}\right]$ is estimated using the PSA simulations of $\inb^{\bm\theta}$ and $\sigma^2_{\bm\phi}$ is the sample variance of $\inb^{\bm\phi_s}$, $s=1,\dots,S$. The EVSI is then calculated from $\eta^{\bm X}_s$ by taking \begin{equation}\mbox{EVSI} = \frac{1}{S} \sum_{s=1}^S \max\left\{0,\eta^{\bm X}_s\right\} - \max\left\{0,\mu\right\}\label{EVSI - calc}.\end{equation}

\section{Calculating the EVSI for different sample sizes}\label{sample-sizes}
A key aspect of designing a trial is to determine the number of patients that should be enrolled. Typically, this is achieved using power calculations that determine the trial sample size by ensuring that a hypothesis test concerning the primary outcome has desirable properties \protect\cite{Chowetal:2017}. To use the EVSI alongside, or as an alternative to, these power calculations requires that the EVSI is estimated across different sample sizes. For example, financial or clinical constraints limit the possible sample sizes for a trial and then the EVSI must be estimated on a grid of $R$ sample sizes between these limits, $\bm N = (N_{\min}=N_1,N_2,\dots,N_R=N_{\max})$.

Using the moment matching method to estimate the EVSI across these sample sizes would require $R \times Q$ posterior updates. This quickly becomes infeasible as $R$, the number of sample sizes considered, increases. Therefore, this section extends the moment matching method to estimate the EVSI across different sample sizes by setting $R=Q$ and only requiring \protect\emph{one} posterior update per sample size. This reduces the required number of posterior updates from $R\times Q$ to $Q$ --- the same as the standard method. 

For this extension, each potential data set $\bm X_q$ is simulated conditional on the vector $(\bm \phi_q,N_q)$, rather than solely on $(\bm\phi_q)$, represented pictorially in Figure \protect\ref{samplingoverN}. Once these datasets $\bm X_1, \dots, \bm X_Q$, with different sample sizes, have been simulated the extended moment matching method proceeds in the standard method; each dataset updates the distribution of the model parameters and then the posterior variance of the incremental net benefit is calculated, giving $\sigma_q^2$ for $q=1,\dots,Q$.

However, to then estimate the EVSI across sample size, we develop an alternative strategy to estimate $\sigma_{\bm X}^2$ from $(\sigma_1^2,\dots,\sigma^2_Q)$. This is because we are now interested in estimating $\sigma_{\bm X}^2(N)$, i.e.~the variance of the expected posterior mean $\mu^{\bm X}(N)$ for different samples of size $N$. In a similar manner to M\"uller and Parmigiani \cite{MullerParmigiani:1995}, we use regression methods to estimate a function $f(N) = \sigma^2_{\bm X}(N)$, using $\sigma^2_q$.  The estimated values $\sigma^2_q$ are generated conditional on a specific future dataset $\bm X_q$, of size $N_q$. This dataset represents one possible future, which is subject to variation due to random sampling and dependence on the specific value of $\bm\phi_q$. In the standard moment matching method, we marginalise out this dependence on the specific $\bm\phi_q$ value by taking the mean of $\sigma^2_q$. However, in this extended setting, we use regression to marginalise out this dependence across the different sample sizes $N_q$. Therefore, we use the values of $\sigma^2_q$, obtained by simulation, to estimate the function $f(N)$ as \[ \sigma^2 - \sigma_q^2= f(N_q) + \varepsilon_q,\] where $\varepsilon_q$ is an error term that captures variation in $\sigma_q^2$ due to using simulated datasets with sample size $N_q$.

\subsection{A Bayesian non-linear model}
To estimate $\sigma^2_{\bm X}(N)$, we define a Bayesian non-linear model which allows us to encode our knowledge about the underlying behaviour of $\sigma_{\bm X}^2(N)$. It also allows us to propagate uncertainty from our estimation procedure into the EVSI estimate so researchers can assess the accuracy of the moment matching procedure. 

Our model is defined by assuming a functional form for the regression function, \[f(N) = \sigma^2_{\bm\phi}\frac{N}{N+h},\] where $\sigma^2_{\bm\phi} = \V\left[\inb^{\bm\phi}\right]$ and $h$ is a parameter to be estimated, and by modelling the residual error as a normal distribution centred on 0, $\varepsilon_q \sim N(0,\sigma^2_{\varepsilon})$ for $q=1,\dots,Q$. This functional form for $f(N)$ reflects the underlying knowledge about $\sigma_{\bm X}^2(N)$. Specifically, we know that $\sigma_{\bm X}^2(N)$ increases as $N$ increases and $f(N)$ is an increasing function. In addition to this, a study with an infinite sample size would be equivalent to gaining \emph{perfect} information about the model parameters $\bm \phi$. This means that $\mu^{\bm X}(N)$ tends to INB$^{\bm\phi}$ as $N \to \infty$, so for an infinite sample size we have $\sigma^2_{\bm X}(\infty) = \sigma^2_{\bm\phi}$. Finally, the functional form of $f(N)$ is the exact relationship between $N$ and $\sigma^2_{\bm X}(N)$ in normal-normal conjugate settings with only one underlying model parameter. So, while realistic health-economic models are not based on normal-normal conjugate models and have multiple model parameters, this function is approximately correct in many settings.

To define the full Bayesian model, priors must be defined for $h$, the regression parameter, and $\sigma^2_{\varepsilon}$, the residual variance. Throughout, we have used a non-central half-Cauchy prior for $\sigma_{\varepsilon}$, as suggested by Gelman \cite{Gelman:2006}, with parameters dependent on the data. This is because the scale of $\sigma_q^2$ changes significantly across health economic models. Therefore, the prior mean is set to the standard deviation of $\sigma^2_q$ divided by 2 and the prior variance to the standard deviation of $\sigma^2_q$. Finally, a data dependent normal prior, truncated at 0, is set for $h$ with the prior mean and variance set to $\frac{N_{\max}}{2}$ and $200\times N_{\max}$ respectively, giving vague priors for $h$ and $\sigma^2_{\varepsilon}$.

\subsection{Calculating the EVSI for different sample sizes}\label{Calculate-EVSI}
To calculate the EVSI, we fit this non-linear model using the estimated posterior variances from the health economic model $\sigma_q^2$. Fitting this model requires Markov Chain Monte Carlo methods to estimate the posterior for $h$ by simulation. The posterior distribution for $\sigma^2_{\bm X}(N)$, irrespective of whether $N$ is in $\bm N$, can be estimated by calculating \[f(N) = \sigma^2_{\bm X}(N) = \sigma_{\bm\phi}^2\frac{N}{N+h}\] for each posterior simulation for $h$. In theory, each simulation from the posterior distribution of $\sigma^2_{\bm X}(N)$ can be used to rescale $\inb^{\bm\phi_s}$, equation (\ref{rescale}), and estimate the EVSI using equation (\ref{EVSI - calc}).

However, this process is computationally intensive if there is a large number of posterior simulations for $h$. Therefore, it is more efficient to calculate the EVSI for a small number of samples from the posterior of $\sigma^2_{\bm X}(N)$. We advise that the posterior of $\sigma^2_{\bm X}(N)$ is summarised by finding credible intervals for a small number of credible levels. The EVSI is then calculated for each estimated variance in this low-dimensional summary. For example, in \S\ref{Examples}, the posteriors for $\sigma^2_{\bm X}(N)$ are summarised using the median and the 75\% and 95\% credible intervals. Note that, the values calculated using this method are not the posterior credible intervals for the EVSI as the relationship between the $\sigma^2_{\bm X}(N)$ and the EVSI is highly non-linear. Code to calculate the EVSI using this method is provided in the supplimentary material.

\begin{table}[!h]
\begin{tabular}{c|p{14cm}}
Notation &Definition\\
\hline
$p(\cdot)$ & The distribution of a random variable.\\
$\bm\theta$&The underlying model parameters for the health economic model.\\
$\bm\phi$& The model parameters that are informed by the trial under consideration.\\
$S$&The number of PSA simulations taken from $p(\bm\theta)$.\\
$\bm X$ & The potential data arising from the future study.\\
INB$^{\bm\theta}$& The incremental net benefit, where uncertainty is induced by the model parameters. \\
INB$^{\bm\phi}$& The conditional expectation of the incremental net benefit conditional on $\bm\phi$.\\
$\mu$ & The expected value of INB$^{\bm\theta}$ across all values of $\bm\theta$.\\
$\sigma^2$ & The variance of INB$^{\bm\theta}$. \\
$\sigma^2_{\bm\phi}$& The variance of INB$^{\bm\phi}$.\\
$\mu^{\bm X}$&  The posterior expectation of INB$^{\bm\theta}$ conditional on the potential datasets.\\
$\mu^{\bm X}(N)$&  The posterior expectation of the incremental net benefit across potential future datasets with a given sample size of $N$.\\
$\sigma^2_{\bm X}$& The variance of $\mu^{\bm X}$.\\
$\sigma^2_{\bm X}(N)$ & The variance of $\mu^{\bm X}(N)$ for potential datasets of a given sample size $N$.\\
$Q$& The number of nested simulations taken to estimate the EVSI using moment matching.\\
$\bm\phi_q$& The $q$-th row of a matrix that contains the quantiles of the PSA simulations for $\bm \phi$ for $q=1,\dots,Q$.\\
$N_q$& The $q$-th element in a vector of sample sizes between the minimum and maximum sample sizes for $q=1,\dots,Q$. \\
$\bm X_q$& The $q$-th simulated potential dataset for $q=1,\dots,Q$. Datasets are simulated conditional on $\bm\phi_q$ only in the standard method and on the pair $(\bm\phi_q,N_q)$ in the extension.\\
$\sigma^2_q$& The posterior variance of the incremental net benefit conditional on the dataset $\bm X_q$ for $q=1,\dots,Q$.\\
$\eta^{\bm X}_s$& The simulated values of INB$^{\bm\phi}$ that have been rescaled to approximate simulations from the distribution of $\mu^{\bm X}$. The subscript $_s$ defines the $s$-th simulated value for $s=1,\dots,S$.\\
$\sigma^2_{\varepsilon}$  & The residual variance for Bayesian non-linear regression used to estimate $\sigma^2_{\bm X}(N)$.
\end{tabular}
\caption{A list of the key notation used to demonstrate the moment matching method and the extension using non-linear regression.}
\label{notation}
\end{table}

\subsection{Practical Considerations}\label{practical}
The first practical consideration is that the function $f(N)$ changes quickly for small sample sizes, meaning that smaller values of $N$ sizes give more information about $h$. Therefore, we suggest that more future samples $\bm X_q$ are generated for small sample sizes. Practically, this is achieved by choosing $N_q$ values that are evenly spaced on the square root scale between $N_{\min}$ and $N_{\max}$. This has a computational advantage as Bayesian updating typically requires more computing time to estimate the posterior distribution conditional on larger datasets. Therefore, choosing $N_q$ in this manner improves fit and reduces computation time.

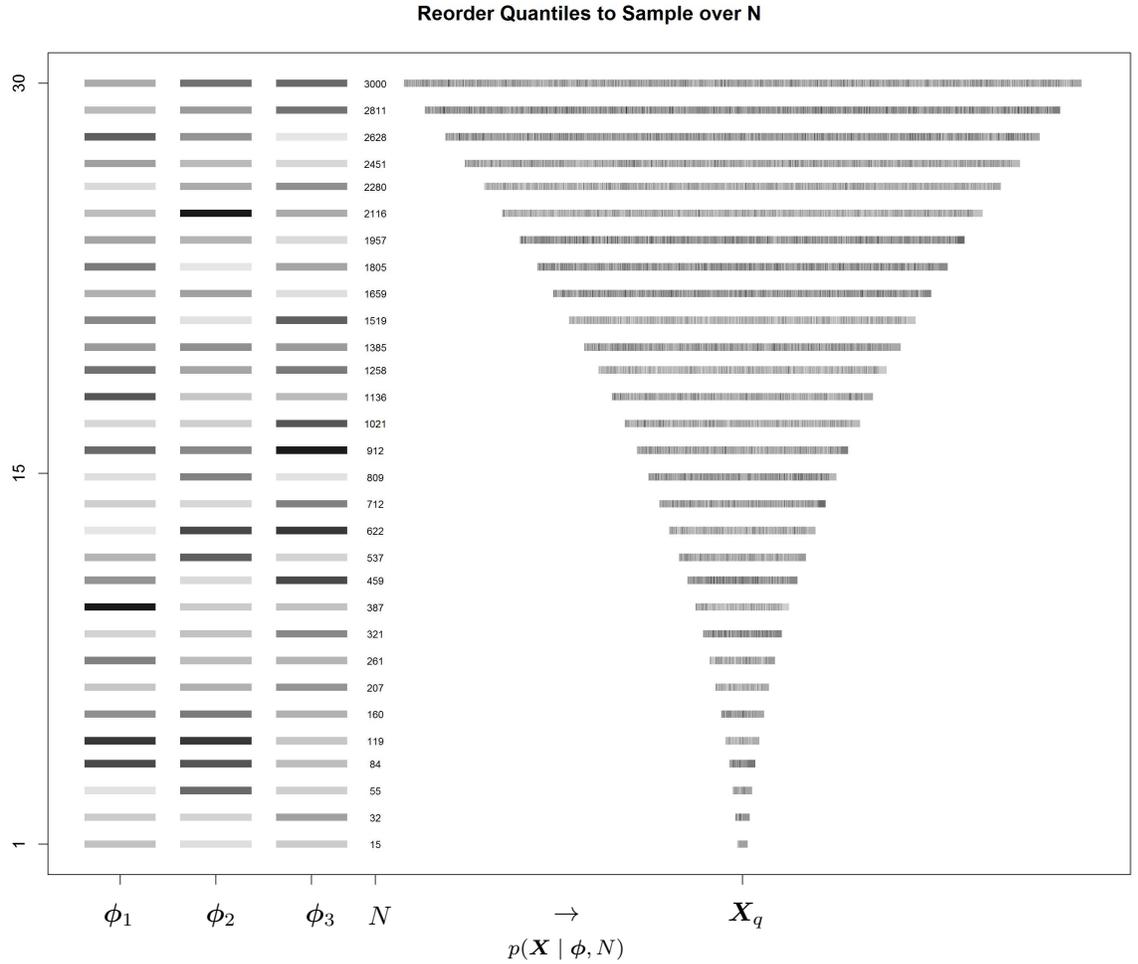
\begin{figure}[!h]
\centering
\begin{tikzpicture}
\node at (0,0) (aa) {\includegraphics[width=16cm]{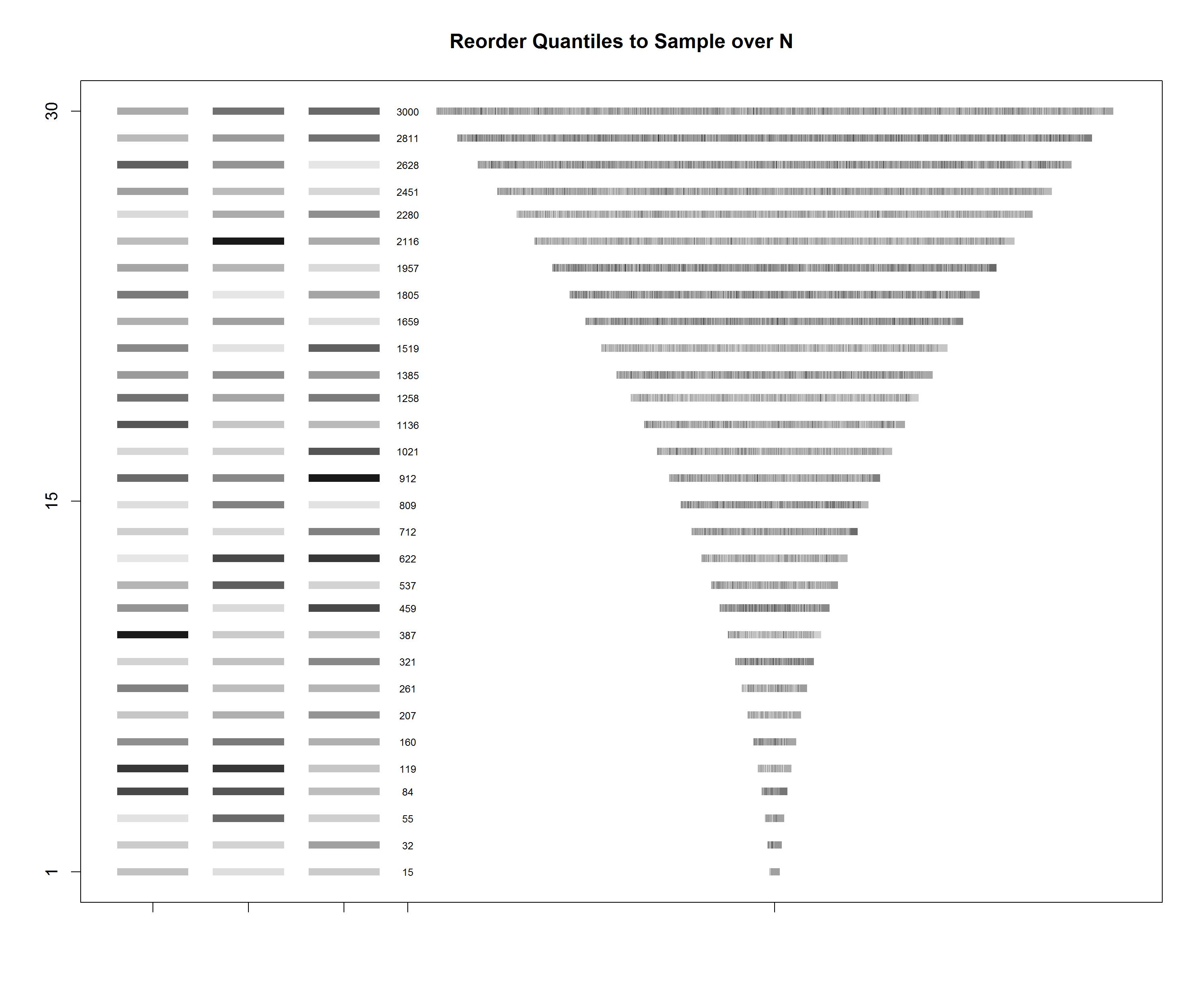}};
\node[below left =-3.4em and -7em of aa,color=black] (bb) {$\bm\phi_1$};
\node[right = 2em of bb,color=black] (cc) {$\bm\phi_2$};
\node[right = 1.9em of cc,color=black] (dd) {$\bm\phi_3$};
\node[right = 0.5em of dd,color=black] (ee) {$N$};

\node[right = 12em of ee,color=black] (ff) {$\bm X_q$};
\node[right = 5.4em of ee,color=black] (gg) {$\rightarrow$};
\node[below = 0em of gg,color=black] (hh) {\footnotesize$p(\bm X \mid \bm\phi,N)$};
\end{tikzpicture}
\caption{A pictorial representation of the sampling procedure for the future samples in the extended moment matching method. As in Figure \ref{MMdescription}, darker lines represent higher values for the parameters.}
\label{samplingoverN}
\end{figure}

Next, this non-linear regression method requires that $\bm\phi_q$ and $N_q$ are uncorrelated. This is because there is normally a relationship between $\sigma^2_{\bm X}(N)$ and $\bm\phi$, through the simulated dataset $\bm X_q$. If $\bm\phi_q$ and $N_q$ are also correlated then this can mask/inflate the relationship between $\sigma^2_{\bm X}(N)$ and $N$. Thus far, $\bm\phi_q$ has been defined as a row containing the sample quantiles for each element of $\bm\phi$ and $N_q$ as increasing values between $N_{\min}$ and $N_{\max}$. This specification naturally induces a correlation between $\bm\phi_q$ and $N_q$ which can lead to biased results. Therefore, it is suggested that each element of $\bm\phi_q$ is randomly reordered separately before generating $X_q$. Reordering each $\bm\phi_q$ vector separately also improves accuracy by inducing a greater variation in the $\bm X_q$ samples. Therefore, the simulation method for $\bm X_q$ is represented in Figure \ref{samplingoverN}, where the quantiles for $\bm\phi$ are reordered separately for each column and then combined with ordered sample sizes, spaced on the square root scales, to generate one potential sample $\bm X_q$ for each sample size.

Finally, as this EVSI estimation method is based on regression, its accuracy can be checked using standard model checking procedures such as residual or qq-plots. Specifically, the residuals should exhibit no clear unmodelled structure, as in standard regression modelling. 

\section{Implementation of the Extended Moment Matching Method} \label{Examples}
This section implements our extended moment matching method to calculate the EVSI across different sample sizes. The supplementary material demonstrates that our method is in line with other EVSI calculation methods \cite{Menzies:2016,Strongetal:2015} using a hypothetical health economic model developed in \cite{BrennanKharroubi:2007}. To demonstrate the power of the extended moment matching method, we determine the optimal trial design for a practical health-economic model that evaluates different treatments for chronic pain \cite{Sullivanetal:2016}. In this analysis, we compare our method with the nested Monte Carlo simulation method to demonstrate accuracy and compare computation time. 

\subsection{A Health Economic Model for Chronic Pain}
Sullivan \textit{et al.} \cite{Sullivanetal:2016} developed a cost-effectiveness model for evaluating treatment options for chronic pain. The model is based on a Markov structure with 10 states, where each state has an associated utility score and cost. The disease progression for chronic pain is as follows: the treatment is administered and a patient can either experience adverse effects of the treatment (AE) or not. The patient can then withdraw from treatment, either due to the AE or otherwise. They can then continue to another treatment option or withdraw completely from treatment. After the second line of treatment, they either have subsequent treatment or discontinue (these are absorbing states).

In the standard model, a patient can either be offered morphine or a innovative treatment in the first line of treatment. If they withdraw and receive a second treatment, they are offered oxycodone, meaning the only difference between the two treatment arms occurs when the first treatment is administered. The innovative treatment is more effective and reduces the probability of AE but is more expensive.

The model uses a willingness-to-pay of $\pounds 20\,000$, the threshold under which treatments are likely to receive a recommendation in the UK \cite{NICE:2013}. For a more in-depth presentation of all the model parameters, see \cite{Sullivanetal:2016}. The PSA is based on gamma distributions for costs and beta distributions for probabilities and utilities. The parameters for these distributions are chosen such that the mean is informed by a literature review and the standard deviation is taken as 10\% of the underlying mean estimate. For illustrative purposes, the per person lifetime EVSI is considered throughout this section. This assumes a discount factor of 0.03 over 15 years of the treatment being available.

\subsection{Analysis for the Chronic Pain Example}
A full VoI analysis, as set out by Tuffaha \textit{et al.} \cite{Tuffahaetal:2016}, is undertaken for the chronic pain model. To begin, 100\,000 simulations are taken from the prior for the model parameters. These are then fed through the Markov model to give 100\,000 simulations for the incremental net benefit. Calculating the EVPPI \cite{Brennanetal:2007,Strongetal:2015,Heathetal:2016} based on these simulations indicated that the most valuable parameters are those relating to the utility of the different health states for the first line of treatment. Specifically, the utility of not having any AE from the treatment and withdrawing from the treatment without experiencing AE. 

In response to this, we designed an experiment to learn about the utility of these two health states, which, if known with certainty, would account for about 79\% of the uncertainty in the decision. Questionnaires are a standard method for determining QALYs and therefore we designed a trial in which questionnaires are sent to $N$ participants. We assume, in a simplistic manner, that a participant is sent the questionnaire if they withdrew from the first treatment without any AE. They are then questioned about both health states, which they can accurately recall. This means that each questionnaire directly updates information about both these key utility parameters.

Responses from these questionnaires will be translated into a utility score for each health state. These utility scores are modelled as independent beta distributions with their means conditional on the utility of interest, the utility of not having AE and withdrawing without AE, and standard deviations equal to 0.3 and 0.31 respectively \cite{Ikenbergetal:2012}. We consider sample sizes between $N_{\min}=10$ and $N_{\max}=150$. It is then assumed that only a proportion of the questionnaires are returned, leading to missingness in the data.

Within this context, two different designs are considered based on the results of a trial investigating whether financial incentives improve the rate of return of questionnaires in clinical trials \cite{Gatesetal:2009}; we assume that the questionnaire could be accompanied by a $\pounds 5$ incentive to complete. This incentive study demonstrated that sending the financial reward improved response rate from 68.7\% to 75.7\% while increasing the cost from $\pounds 4.64$ to $\pounds 9.35$. The difference in cost is not equal to the incentive because non-respondents are chased up by telephone, which costs staff time and other resources and sending the reward reduces the number of chase-up calls needed.

Therefore, the EVSI will be used to answer two key questions regarding the design of this study; firstly, should the incentive be used? and secondly, how many questionnaires should be sent? The EVSI is estimated for both response rates using $Q=50$ with 10\,000 posterior simulations fed through the Markov model and used to estimate $\sigma^2_q$ for each $q=1,\dots,Q$. The EVSI results using the moment matching method are compared with nested Monte Carlo simulations for sample sizes 10, 25, 50, 100 and 150. These nested MC estimators were calculated using 100\,000 PSA simulations to generate the future samples and then 100\,000 posterior simulations per PSA simulation. So, 1 billion simulations were used for each sample size to estimate the EVSI using nested Monte Carlo simulations as opposed to 500\,000 simulations for our novel method.

\subsection{Results for the Chronic Pain Model} \label{pain-results}
\begin{figure}[!h]
\begin{center}
\includegraphics[width=12cm]{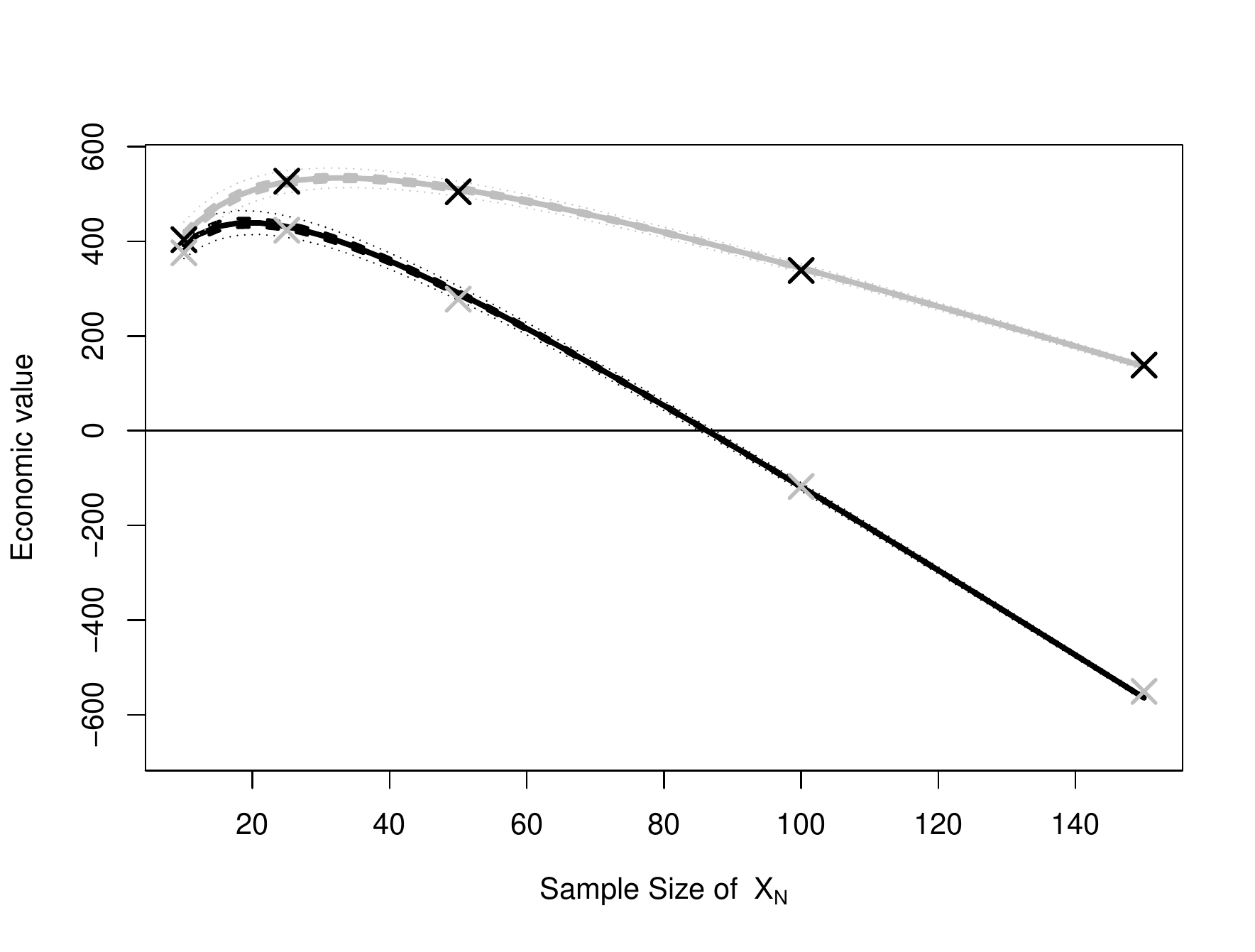} 
\caption{The economic value (difference between EVSI and the cost of undertaking the study) of the two alternative questionnaire collection strategies for different sample sizes in the chronic pain example. The grey curves indicate the potential value of the study without incentives and the black curves the value of the study with incentives estimated using the moment matching method. The black crosses represent the no-incentive Monte Carlo simulation estimates and the grey crosses represent the same estimates for the incentive study. }
\label{Pain}
\end{center}
\end{figure}

Figure \ref{Pain} displays the net economic value for both designs; the grey curves are the economic value of the study with no incentive and the black curves are the value of the incentive study from the moment matching method.  The economic value of a study is the difference between the EVSI and its cost, from \cite{Gatesetal:2009}. Clearly, the no incentive study is economically more valuable. This is because the reduced missingness in the questionnaire responses is not valuable enough to warrant the cost of the incentive. Note, however, that the dominance of the no incentive study is uncertain for very small sample sizes as the EVSI ``distributions'' overlap. 

Figure \ref{Pain} also shows the nested Monte Carlo estimates for the economic study value; the black crosses give the value for the no incentive study and the grey crosses for the incentive study. Using these estimates, we demonstrate that the moment matching method with non-linear regression is in line with the nested Monte Carlo simulation. This means that, in this example, our novel EVSI calculation method is accurate, despite the non-linear and non-Gaussian model structure of the health economic model, and significantly reduces the required number of simulations for this analysis.

Analysing the economic value for different sample sizes determines that the optimal sample size in terms of economic benefit is 16 for the incentive study and 27 for the no incentive study. Therefore, overall, the optimal study is a no incentive study with 27 participants; the information gained from any additional patients is not worth the cost of sending the questionnaire and chasing up the non-respondents.

The total computational time to calculate the EVSI for both studies using our novel calculation method is 331 seconds or around 5 and a half minutes on a standard desktop computer with a Intel i7 Pro processor using \texttt{R} version 3.2.1. This includes fitting the regression model to calculate the EVSI across different sample sizes for both study designs, which is around 5 seconds per study. On the other hand, the nested Monte Carlo simulations required between between 7 and 68 days for each EVSI estimate, with a total computation time of around 311 days for all 10 estimates. 

\section{Discussion}
The EVSI has frequently been touted as a method for study design, including determining the optimal study size. However, its application in this area has been limited by the computational time required to obtain EVSI estimates across different designs. Several approximation methods have reduced the computational time per EVSI estimate to allow for this type of analysis. However, even with these calculation methods, the computational time increases linearly with the number of sample sizes considered. 

Therefore, in this paper we present an extension to the moment matching method for EVSI calculation \cite{Heathetal:2017b} that performs the analysis over different sample sizes with a minimal fixed additional computational cost. This implies that the cost of estimating the EVSI for a single design is approximately equal to the evaluation of the EVSI across different sample sizes. This method augments the moment matching method with Bayesian non-linear regression. This estimates the EVSI by calculating the posterior variance of the incremental net benefit for around 50 different sample sizes. In general, the posterior variance is easy to calculate provided a sampling distribution can be defined for the data and a Bayesian model can be specified for the model parameters.

The augmented moment matching method determined the optimal study design to update the information underpinning a real-life health economic model that evaluates interventions for the treatment of chronic pain. We also demonstrated that our novel method accurately estimates the EVSI by comparing with nested Monte Carlo simulations. 

 This method allows researchers to determine the optimal sample size for their trials using EVSI. The minimal computational time also allows them to compare a number of other design features. Finally, this method is implemented in the \texttt{R} package \texttt{EVSI} \cite{EVSI:2017} making this method relatively simple to implement for general health economic models. The \texttt{EVSI} package also extends the methodology to calculate the EVSI across different net benefit functions at a minimal extra cost. Finally, the package also includes a suite of graphics and an online tool so EVSI results can be presented to key stakeholders \cite{Heathetal:2018c}.

\section*{Acknowledgments}
 Financial support for this study was provided in part by grants from Engineering and Physical Sciences Research Council (EPSRC) [Anna Heath] and Mapi [Dr. Gianluca Baio]. The funding agreement ensured the authors' independence in designing the study, interpreting the data, writing, and publishing the report. The authors would also like to thank Nick Menzies for providing the simulations for the BK example included in the supplementary material.

\bibliographystyle{plain}
\bibliography{bib}

\begin{thebibliography}{10}

\bibitem{Adesetal:2004}
A.~Ades, G.~Lu, and K.~Claxton.
\newblock {Expected Value of Sample Information Calculations in Medical
  Decision Modeling}.
\newblock {\em Medical Decision Making}, 24:207--227, 2004.

\bibitem{AndronisBarton:2016}
L.~Andronis and P.~Barton.
\newblock Adjusting estimates of the expected value of information for
  implementation: theoretical framework and practical application.
\newblock {\em Medical Decision Making}, 36(3):296--307, 2016.

\bibitem{BCEA:R:2016}
G.~Baio, A.~Berardi, and A.~Heath.
\newblock {\em BCEA: Bayesian Cost Effectiveness Analysis}, 2016.
\newblock R package version 2.2-3.

\bibitem{BCEA:2017}
G.~Baio, A.~Berardi, and A.~Heath.
\newblock {\em {Bayesian Cost Effectiveness Analysis with the \texttt{R}
  package \texttt{BCEA}}}.
\newblock {Springer}, 2017.

\bibitem{BaioDawid:2011}
G.~Baio and P.~Dawid.
\newblock {Probabilistic sensitivity analysis in health economics}.
\newblock {\em Statistical methods in medical research}, Sep 2011.

\bibitem{BrennanKharroubi:2007}
A.~Brennan and S.~Kharroubi.
\newblock Efficient computation of partial expected value of sample information
  using bayesian approximation.
\newblock {\em Journal of health economics}, 26(1):122--148, 2007.

\bibitem{BrennanKharroubi:2007b}
A.~Brennan and S.~Kharroubi.
\newblock Expected value of sample information for weibull survival data.
\newblock {\em Health economics}, 16(11):1205--1225, 2007.

\bibitem{Brennanetal:2007}
A.~Brennan, S.~Kharroubi, A.~O'Hagan, and J.~Chilcott.
\newblock {Calculating Partial Expected Value of Perfect Information via Monte
  Carlo Sampling Algorithms}.
\newblock {\em Medical Decision Making}, 27:448--470, 2007.

\bibitem{CADTH:2006}
{Canadian Agency for Drugs and Technologies in Health}.
\newblock {Guidelines for the economic evaluation of health technologies:
  Canada [3rd Edition].}, 2006.

\bibitem{Chowetal:2017}
S.~Chow, J.~Shao, H.~Wang, and Y.~Lokhnygina.
\newblock {\em Sample size calculations in clinical research}.
\newblock Chapman and Hall/CRC, 2017.

\bibitem{Claxtonetal:2005}
K.~Claxton, M.~Sculpher, C.~McCabe, A.~Briggs, R.~Akehurst, M.~Buxton,
  J.~Brazier, and A.~O'Hagan.
\newblock {Probabilistic sensitivity analysis for NICE technology assessment:
  not an optional extra}.
\newblock {\em Health Economics}, 14:339--347, 2005.

\bibitem{Australia:2008}
{Department of Health and Ageing}.
\newblock {Guidelines for preparing submissions to the Pharmaceutical Benefits
  Advisory Committee: Version 4.3}, 2008.

\bibitem{eunethta:2014}
EUnetHTA.
\newblock {Methods for health economic evaluations: A guideline based on
  current practices in Europe - second draft}, 29th September 2014.

\bibitem{Gatesetal:2009}
S.~Gates, M.~Williams, E.~Withers, E.~Williamson, S.~Mt-Isa, and S.~Lamb.
\newblock Does a monetary incentive improve the response to a postal
  questionnaire in a randomised controlled trial? the mint incentive study.
\newblock {\em Trials}, 10(1):44, 2009.

\bibitem{Gelman:2006}
A.~Gelman.
\newblock Prior distributions for variance parameters in hierarchical models.
\newblock {\em Bayesian analysis}, 1(3):515--534, 2006.

\bibitem{EVSI:2017}
A.~Heath and G.~Baio.
\newblock Evsi: A suite of functions for the calculation and presentation of
  the evsi.
\newblock {\em GitHub repository}, 2017.

\bibitem{Heathetal:2018}
A.~Heath and G.~Baio.
\newblock An efficient calculation method for the expected value of sample
  information: Can we do it? yes, we can.
\newblock {\em Value in Health}, -(-):--, 2018.

\bibitem{Heathetal:2018c}
A.~Heath, G.~Baio, and R.~Hunter.
\newblock Development of a new software tool to compute the expected value of
  sample information: An application to the homehealth intervention.
\newblock {\em HESG2018}, 2018.

\bibitem{Heathetal:2016}
A.~Heath, I.~Manolopoulou, and G.~Baio.
\newblock Estimating the expected value of partial perfect information in
  health economic evaluations using integrated nested laplace approximation.
\newblock {\em Statistics in medicine}, 2016.

\bibitem{Heathetal:2017}
A.~Heath, I.~Manolopoulou, and G.~Baio.
\newblock A review of methods for analysis of the expected value of
  information.
\newblock {\em Medical Decision Making}, 37(7):747--758, 2017.

\bibitem{Heathetal:2017b}
A.~Heath, I.~Manolopoulou, and G.~Baio.
\newblock Efficient monte carlo estimation of the expected value of sample
  information using moment matching.
\newblock {\em Medical Decision Making}, 38(2):163--173, 2018.

\bibitem{Howard:1966}
R.~Howard.
\newblock {Information Value Theory}.
\newblock In {\em {IEEE Transactions on System Science and Cybernetics}}, (1)
  22-26, 1966. SCC-2.

\bibitem{Ikenbergetal:2012}
R.~Ikenberg, N.~Hertel, Andrew M., M.~Obradovic, G.~Baxter, P.~Conway, and
  H.~Liedgens.
\newblock Cost-effectiveness of tapentadol prolonged release compared with
  oxycodone controlled release in the uk in patients with severe non-malignant
  chronic pain who failed 1st line treatment with morphine.
\newblock {\em Journal of medical economics}, 15(4):724--736, 2012.

\bibitem{JalalAlarid:2018}
H.~Jalal and F.~Alarid-Escudero.
\newblock A gaussian approximation approach for value of information analysis.
\newblock {\em Medical Decision Making}, 38(2):174--188, 2018.

\bibitem{Jalaletal:2015}
H.~Jalal, J.~Goldhaber-Fiebert, and K.~Kuntz.
\newblock Computing expected value of partial sample information from
  probabilistic sensitivity analysis using linear regression metamodeling.
\newblock {\em Medical Decision Making}, 35(5):584--595, 2015.

\bibitem{McKennaClaxton:2011}
C.~McKenna and K.~Claxton.
\newblock Addressing adoption and research design decisions simultaneously: the
  role of value of sample information analysis.
\newblock {\em Medical Decision Making}, 31(6):853--865, 2011.

\bibitem{Menzies:2016}
N.~Menzies.
\newblock An efficient estimator for the expected value of sample information.
\newblock {\em Medical Decision Making}, 36(3):308--320, 2016.

\bibitem{Migueletal:2016}
A.~Miquel-Cases, V.~Ret{\`e}l, W.~van Harten, and L.~Steuten.
\newblock Decisions on further research for predictive biomarkers of high-dose
  alkylating chemotherapy in triple-negative breast cancer: a value of
  information analysis.
\newblock {\em Value in health}, 19(4):419--430, 2016.

\bibitem{MullerParmigiani:1995}
P.~M{\"u}ller and G.~Parmigiani.
\newblock Optimal design via curve fitting of monte carlo experiments.
\newblock {\em Journal of the American Statistical Association},
  90(432):1322--1330, 1995.

\bibitem{NICE:2013}
{National Institute of Health and Care Excellence}.
\newblock {Guide to the methods of technology appraisal 2013}, {2013}.

\bibitem{Pandoretal:2015}
A.~Pandor, P.~Thokala, S.~Goodacre, E.~Poku, J.~Stevens, S.~Ren, A.~Cantrell,
  G.~Perkins, M.~Ward, and J.~Penn-Ashman.
\newblock Pre-hospital non-invasive ventilation for acute respiratory failure:
  a systematic review and cost-effectiveness evaluation.
\newblock {\em Health Technology Assessment}, 19(42):1--102, 2015.

\bibitem{RaiffaSchlaifer:1961}
H.~Raiffa and H.~Schlaifer.
\newblock {\em {Applied Statistical Decision Theory}}.
\newblock Harvard University Press, Boston, MA, 1961.

\bibitem{Steutenetal:2013}
L.~Steuten, G.~van~de Wetering, K.~Groothuis-Oudshoorn, and V.~Ret{\`e}l.
\newblock {A systematic and critical review of the evolving methods and
  applications of value of information in academia and practice}.
\newblock {\em Pharmacoeconomics}, 31(1):25--48, 2013.

\bibitem{Strong:2012:Code}
M.~Strong.
\newblock {Partial EVPPI Functions}.
\newblock
  \url{http://www.shef.ac.uk/polopoly_fs/1.305039!/file/R_functions.txt}, 2012.

\bibitem{Strongetal:2015}
M.~Strong, J.~Oakley, A.~Brennan, and P.~Breeze.
\newblock Estimating the expected value of sample information using the
  probabilistic sensitivity analysis sample a fast nonparametric
  regression-based method.
\newblock {\em Medical Decision Making}, page 0272989X15575286, 2015.

\bibitem{Sullivanetal:2016}
W.~Sullivan, M.~Hirst, S.~Beard, D.~Gladwell, F.~Fagnani, J.~Bastida,
  Cl~Phillips, and W.~Dunlop.
\newblock Economic evaluation in chronic pain: a systematic review and de novo
  flexible economic model.
\newblock {\em The European Journal of Health Economics}, 17(6):755--770, 2016.

\bibitem{Tuffahaetal:2016}
H.~Tuffaha, L.~Gordon, and P.~Scuffham.
\newblock Value of information analysis informing adoption and research
  decisions in a portfolio of health care interventions.
\newblock {\em MDM Policy \& Practice}, 1(1):2381468316642238, 2016.

\bibitem{Weltonetal:2011}
N.~Welton, J.~Madan, and A.~Ades.
\newblock Are head-to-head trials of biologics needed? the role of value of
  information methods in arthritis research.
\newblock {\em Rheumatology}, 50(suppl\_4):iv19--iv25, 2011.

\bibitem{Weltonetal:2014}
N.~Welton, J.~Madan, D.~Caldwell, T.~Peters, and A.~Ades.
\newblock Expected value of sample information for multi-arm cluster randomized
  trials with binary outcomes.
\newblock {\em Medical Decision Making}, 34(3):352--365, 2014.

\bibitem{WillanPinto:2005}
A.~Willan and E.~Pinto.
\newblock The value of information and optimal clinical trial design.
\newblock {\em Statistics in medicine}, 24(12):1791--1806, 2005.

\end{thebibliography}

\appendix
\section{Multi-Decision Models}
In general, there is little theoretical difference between the moment matching method for multi-decision models and the moment matching method in the dual-decision setting that has been presented in the paper. The method still proceeds by calculating the posterior variance of the incremental net benefit across different potential simulated datasets. These calculated variances are then used to rescale the expectation of INB$^{\bm\phi}$. However, as we have more than two decisions, the incremental net benefit is defined in a slightly different manner.

In the multi-decision setting, we choose a reference treatment, say $T$, and then calculate the incremental net benefit of all the treatments with respect to this reference treatment, i.e.~\[\mbox{INB}^{\bm\theta}_t=\mbox{NB}_t^{\bm\theta}-\mbox{NB}_T^{\bm\theta}.\] Therefore, the incremental net benefit INB$^{\bm\theta}$ can be thought of as a multivariate random vector: 
\begin{equation} 
\inb^{\bm\theta} = \left(
\begin{array}{c}
\inb^{\bm\theta}_1\\
\inb^{\bm\theta}_2\\
\vdots\\
\inb^{\bm\theta}_{T-1}
\end{array}\right).
\end{equation}
This means that the variance of INB$^{\bm\theta}$ is a variance-covariance matrix rather than a scalar value and so we denote it $\Sigma_{\bm\theta}$. If the EVSI calculation is being performed in \texttt{R}, then the variance-covariance matrix is computed when using the \texttt{var()} function so this extension adds no complexity to the moment matching procedure.

To find INB$^{\bm\phi}$ using non-parametric regression, a regression curve should be fitted for incremental net benefit to give a multivariate vector:
\begin{equation} 
\inb^{\bm\phi} = \left(
\begin{array}{c}
\inb^{\bm\phi}_1\\
\inb^{\bm\phi}_2\\
\vdots\\
\inb^{\bm\phi}_{T-1}
\end{array}\right),
\end{equation}
with a variance covariance matrix denoted $\Sigma_{\bm\phi}$. The standard moment matching method would proceed as normal, i.e.~for each potential sample $\bm X_q$, we would calculate the variance of INB$^{\bm\theta}$, the only difference in the multi-decision setting is that the variance is a variance-covariance matrix which we denote $\Sigma_q$.

In the multi-decision setting, as before, we are aiming to estimate the distribution of $\mu^{\bm X} = \mbox{E}_{\bm\theta\mid\bm X} \left(\mbox{INB}^{\bm\theta}\right)$. However, this is now a multivariate distribution with a variance-covariance matrix, denoted $\Sigma_{\bm X}$.  Each element of this matrix, $\sigma_{\bm X}^{ij}$ for $i=1,\dots,T-1$ and $j=1,\dots,T-1$, is calculated in a similar manner as the dual decision setting, \[\sigma^{ij}_{\bm X} = \sigma^{ij}_{\bm\theta} - \frac{1}{Q} \sum_{q=1}^Q \sigma^{ij}_{q},\] where $\sigma^{ij}_{\bm\theta}$ is the element of the $i$-th row and $j$-th column of the $\Sigma_{\bm\theta}$ matrix and $\sigma^{ij}_{q}$ is the same element of the $\Sigma_q$ matrices. This is therefore, the same formula as the dual-decision setting but separately for each element of the variance-covariance matrix. 

To rescale the distribution of INB$^{\bm\phi}$, we use the same formula as in the standard method, but, rather than dividing by the standard deviation we must multiply by the inverse matrix square root. Matrix square roots and inverses are well-defined and can easily be found in \texttt{R} using the \texttt{expm} package. Therefore, to rescale the simulated PSA vectors for INB$^{\bm\phi}$, denoted INB$^{\bm\phi_s}$, $s=1,\dots,S$, we use the following formula:
\[\eta^{\bm X}_s = (\mbox{INB}^{\bm\phi_s} - \mu)\Sigma_{\bm\phi}^{-\frac{1}{2}}\Sigma_{\bm X}^{\frac{1}{2}} +\mu.\]The EVSI is then calculated by taking the row-wise maximum of each of the $\eta^{\bm X}_s$ vectors and 0 and then taking the mean of these maximums.

Using non-linear regression to calculate the EVSI in multi-decision problems involves an extension to the standard method. The non-linear model defined in the main paper is a scalar function and we must estimate a variance-covariance matrix across different sample sizes. Therefore, we must extend this regression model to a matrix function. In practice, we suggest that the non-linear regression function is extended by fitting the non-linear model from main paper \[f(N) = \sigma^2_{\bm\phi}\frac{N}{N+h}\] separately for each unique element of the variance-covariance matrix. Essentially, this involves estimating the element in the $i$-th row and $j$-th column of the variance matrix $\Sigma_{\bm X}(N)$ for a sample size $N$, denoted $\sigma^{ij}_{\bm X}(N)$ as \[f^{ij}(N) = \sigma^{ij}_{\bm X}(N) = \sigma_{\bm\phi}^{ij} \frac{N}{N+h^{ij}},\] where $\sigma_{\bm\phi}^{ij}$ is the $i$-th, $j$-th element of the $\Sigma_{\bm\phi}$ matrix. It is possible to demonstrate in a decision model with three treatment options that these functions approximately estimate the variance of both incremental net benefits and their covariance in normal-normal conjugate settings. 

As the covariance matrix is symmetric, $f^{ij}(N)=f^{ji}(N)$ and so we fit \[\frac{(T-1)T}{2}\] regression models separately to calculate the EVSI across different sample sizes, where $T$ is the number of possible interventions. Each curve will produce an estimate for the $h^{ij}$ parameter and these distributions can  be combined to find the posterior distribution of the variance-covariance matrix for $\Sigma_{\bm X}(N)$ for each sample size $N$ under consideration.  

Finally, recall that the distribution of variance-covariance matrix for $\sigma^2_{\bm X}(N)$ was approximated by a low-dimensional summary of the distribution of the posterior distribution for $h$ in the main paper. Clearly, this is more challenging when multiple curves are being fitted. Nonetheless, we suggest that $h^{ij}$ posterior is summarised by finding a low-dimensional summary and then these are then used to create a small number of variance-covariance matrices. These matrices are used to rescale are then used to rescale INB$^{\bm\phi}$ using the formula above. The EVSI is finally calculated with the same method in the standard moment matching method. This method does not give credible intervals for the EVSI but computationally efficient way to calculate a low-dimensional summary of the possible EVSI values.

\section{Calculating the EVSI using Moment Matching for the Brennan and Kharroubi Example}
This model has frequently been used to assess calculation methods for the EVSI. It was first developed by Brennan and Kharroubi \cite{BrennanKharroubi:2007} and modified by Menzies \cite{Menzies:2016} to compare two treatments used to treat a hypothetical disease. For each drug, a patient can respond to the treatment, experience side effects or visit hospital for a certain length of time. A utility value is assigned to each of these possible outcomes and costs are associated with the drugs and hospital stays. 

All the parameters are assumed to be normal with the mean and standard deviation given in Table \ref{parameters}. The studies are also assumed to have normal distributions, with the standard deviations given in Table \ref{parameters}. In this example, it is assumed that $\theta_5, \theta_7,\theta_{14}$ and $\theta_{16}$ are correlated with correlation coefficient 0.6 and the parameters $\theta_6$ and $\theta_{15}$ are also correlated with a correlation coefficient 0.6 and independent of the other set of parameters.

\begin{table}
\begin{tabular}{|l|c|c|c|c|c|c|}
\hline
& \multicolumn{2}{c|}{Mean} & \multicolumn{2}{c|}{Standard Deviation (SD)} & \multicolumn{2}{c|}{Data SD}\\
\cline{2-7}
Parameter & $t=0$&$t=0$&$t=0$&$t=1$&$t=0$&$t=1$\\
\hline
Drug Cost ($\theta_{1},\theta_{11}$) &$\$ 10\,000$ &$\$15\,000$&$\$10$&$\$10$& -&-\\
Probability of Hospitalisation ($\theta_2,\theta_{12}$)&0.1&0.08&0.02&0.02&-&-\\
Days in Hospital ($\theta_3,\theta_{13}$)&5.2&6.1&1&1&-&-\\
Hospital Cost per Day ($\theta_4$)&$\$4\,000$&$\$4\,000$&$\$2\,000$&$\$2\,000$& - & - \\
Probability of Responding ($\theta_5,\theta_{14}$)&0.7&0.8&0.1&0.1&0.2&0.2\\
Utility Change due to Response ($\theta_6,\theta_{15}$)&0.3&0.3&0.1&0.05&0.2&0.2\\
Duration of Response (years) ($\theta_7,\theta_{16}$)&3&3&0.5&1&1&2\\
Probability of Side Effects ($\theta_8,\theta_{17}$) &0.25&0.2&0.1&0.05&-&-\\
Utility Change due to Side Effects ($\theta_9,\theta_{18}$)&-0.1&-0.1&0.02&0.02&-&-\\
Duration of Side Effects (years) ($\theta_{10},\theta_{19}$)&0.5&0.5&0.2&0.2&-&-\\
\hline
\end{tabular}
\caption{The parameters for the Brennan and Kharroubi example. The mean and standard deviations for the distributions of the parameters is also given, along with the standard deviation of the data collection exercise aimed at reducing uncertainty in that parameter}
\label{parameters}
\end{table}

The net benefits for each treatment are calculated as a deterministic function of these parameters \[\mbox{NB}_1 = \lambda(\theta_5\theta_6\theta_7 + \theta_8\theta_9\theta_{10}) - (\theta_1 +\theta_2\theta_3\theta_4),\]\[\mbox{NB}_2 = \lambda(\theta_{14}\theta_{15}\theta_{16} +\theta_{17}\theta_{18}\theta_{19}) - (\theta_{11}+\theta_{12}\theta_{13}\theta_4),\]
with  $\lambda=\$100\,000$. Five alternative data collection exercises are proposed by Menzies and are also considered in this exploration: \begin{enumerate}
\item A clinical trial collecting information on the probability that a patient responds to the two treatment options which informs parameters $\theta_{5}$ and $\theta_{14}$.
\item  A study looking at the utility improvement for responding to the different treatments which informs parameters $\theta_6$ and $\theta_{15}$.
\item A study investigating the duration of response to the therapy (for those who do respond), informing parameters $\theta_7$ and $\theta_{16}$.
\item A study combining the first two studies, i.e.~informing $\theta_5,\theta_6,\theta_{14}$ and $\theta_{15}$.
\item A study combining all the previous studies and therefore informing $\theta_5,\theta_6,\theta_7,\theta_{14},\theta_{15}$ and $\theta_{16}$.
\end{enumerate}

\subsection{Analysis for the BK example}
To estimate the EVSI for different sample sizes using the moment matching method, the PSA distribution for the incremental net benefit is estimated using 1 million simulations from the parameter distributions. This implies that $\sigma^2$ and $\mu$, the variance and mean of the incremental net benefit respectively, are estimated using this full sample. $\inb^{\bm\phi}$ are also found using these 1 million simulations, expect for exercise 5 which is based on 6 underlying parameters meaning that the computational demands of estimating $\inb^{\bm\phi}$ was too high. These fitted values are, therefore, based on 20\,000 simulations and obtained using the \texttt{R} package \texttt{BCEA} \cite{BCEA:R:2016,BCEA:2017}. 

In line with Menzies \cite{Menzies:2016}, sample sizes between $N_{\min}=10$ and $N_{\max}=200$ are considered for each of the different exercises outlined above. Throughout the analysis, we set $Q=50$ which implies that 10\,000 simulations are taken from 50 different posterior distributions to calculate the variance of the posterior incremental net benefit for 50 different sample sizes. The distribution for the EVSI is then determined using the method described in \S\ref{Calculate-EVSI} in the main paper.

The results determined using our method are compared with the nested Monte Carlo approach for calculating the EVSI and Menzies' approach which also reweights the PSA simulations for the INB but with an alternative method. These results are taken directly from Menzies \cite{Menzies:2016} and are the most accurate estimates available. The conventional approach required 1 billion model evaluations per sample size compared with 500\,000 model evaluations for the moment matching method to estimate the EVSI across the different sample sizes. 

\subsection{Results for the BK example}\label{BK-results}
\begin{figure}[!h]
\begin{center}
\includegraphics[width=8cm]{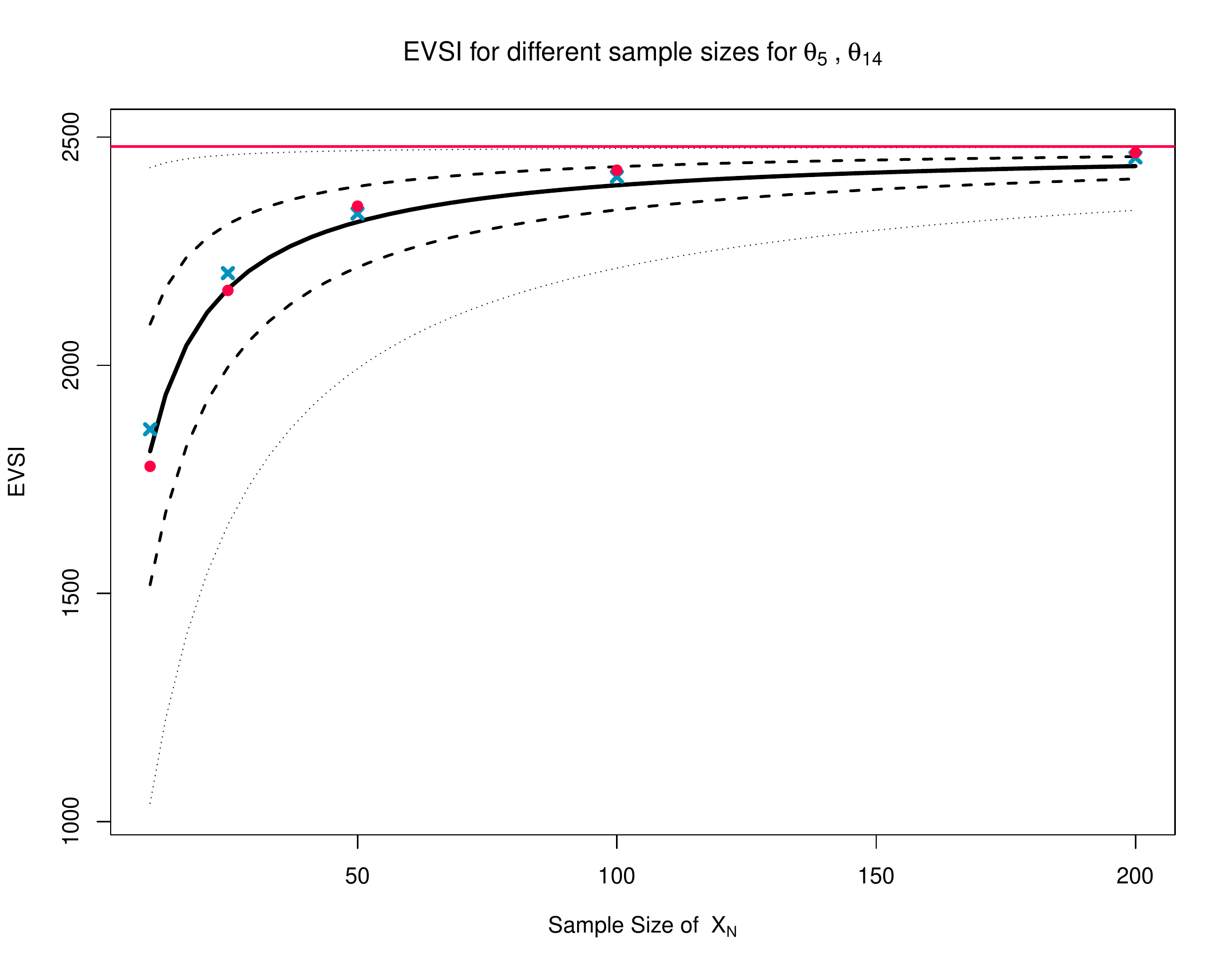}
\includegraphics[width=8cm]{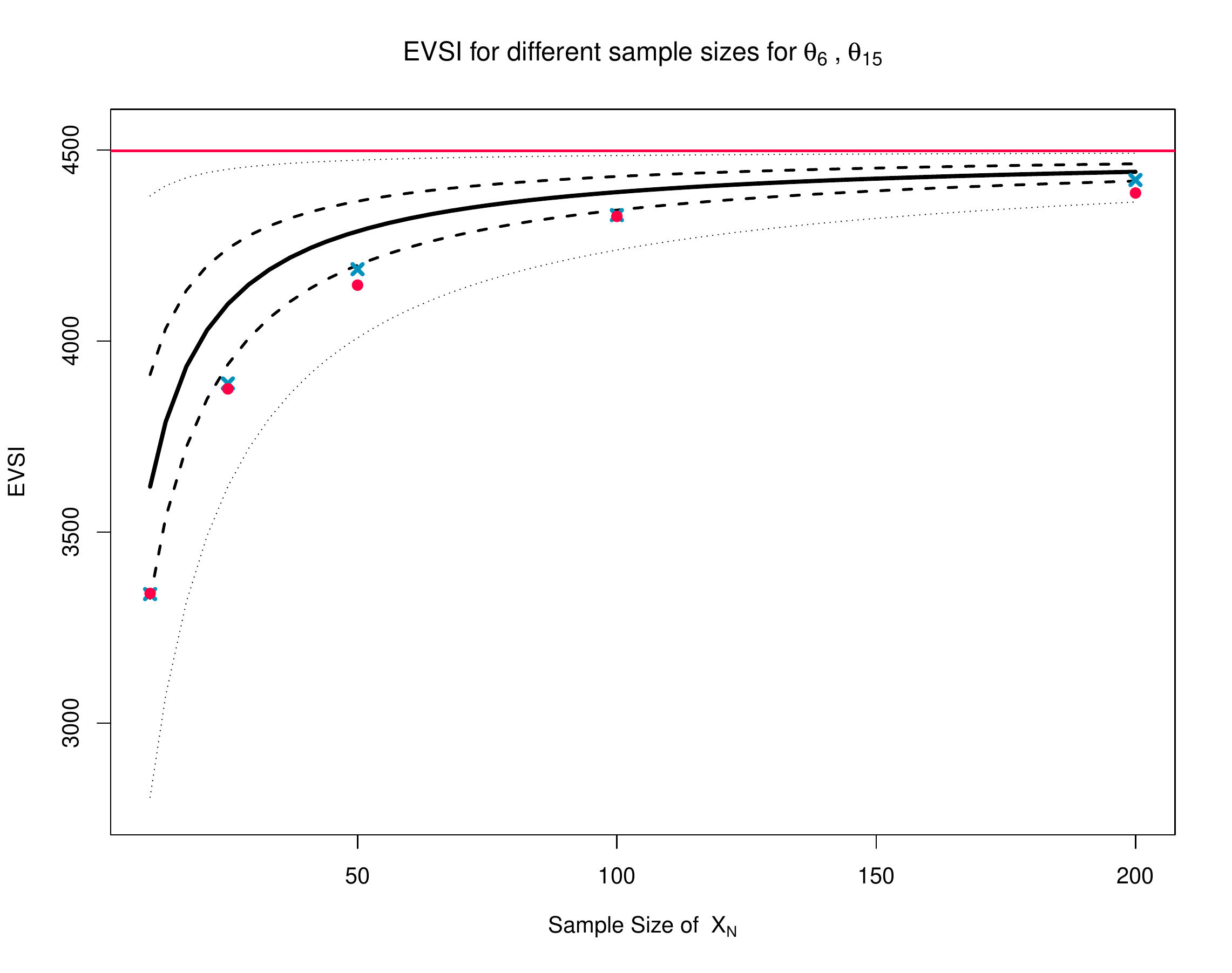}
\includegraphics[width=8cm]{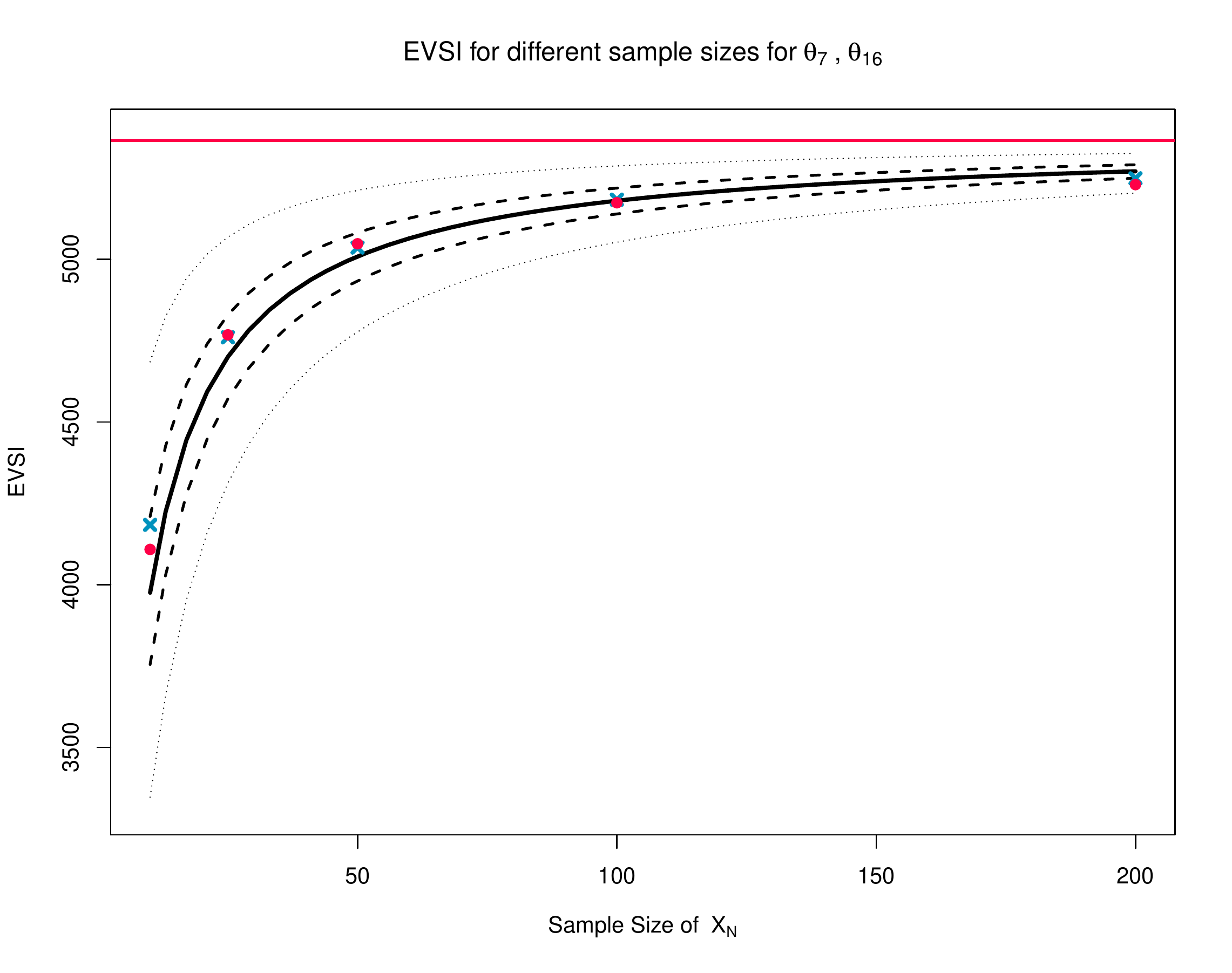}
\includegraphics[width=8cm]{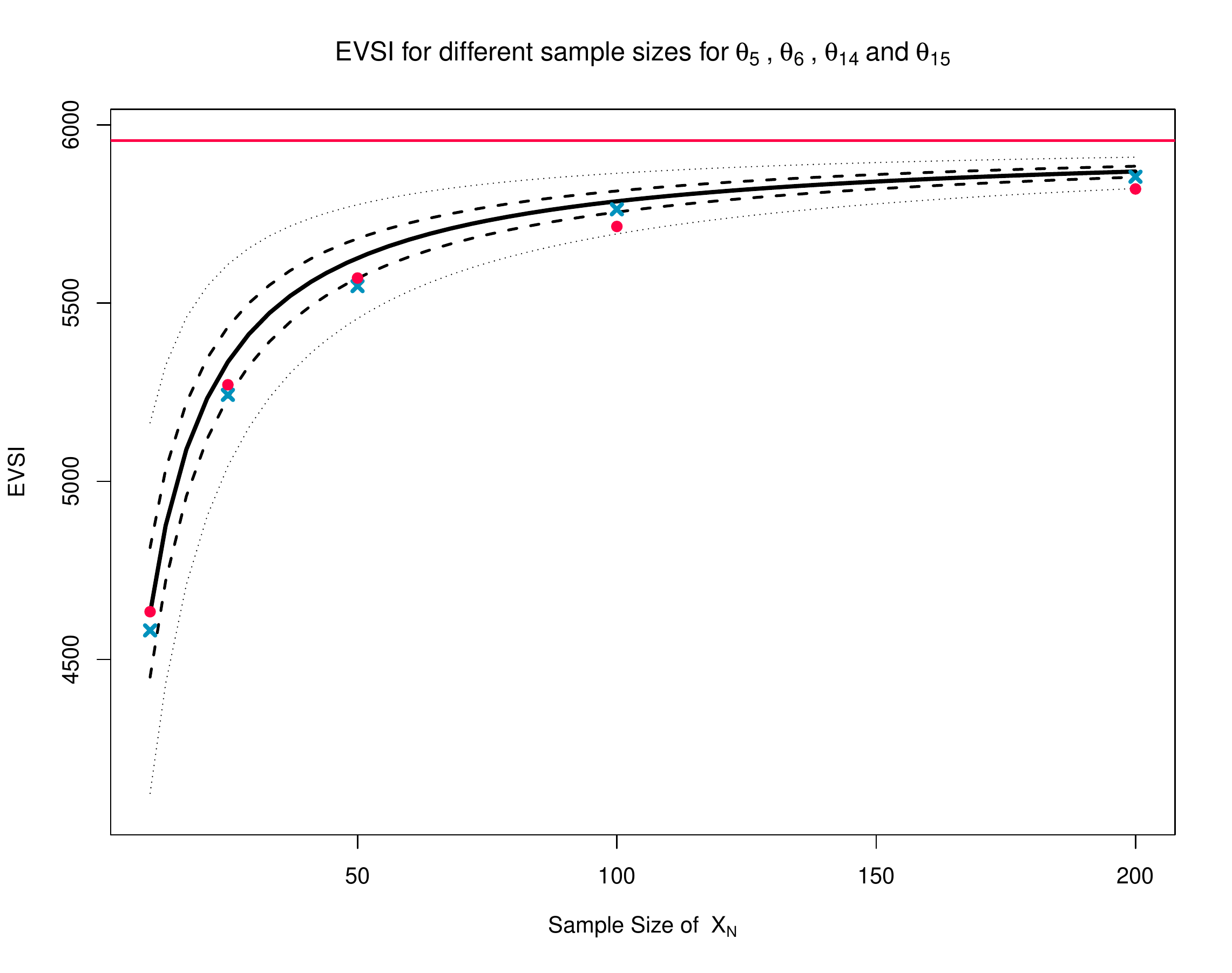}
\includegraphics[width=8cm]{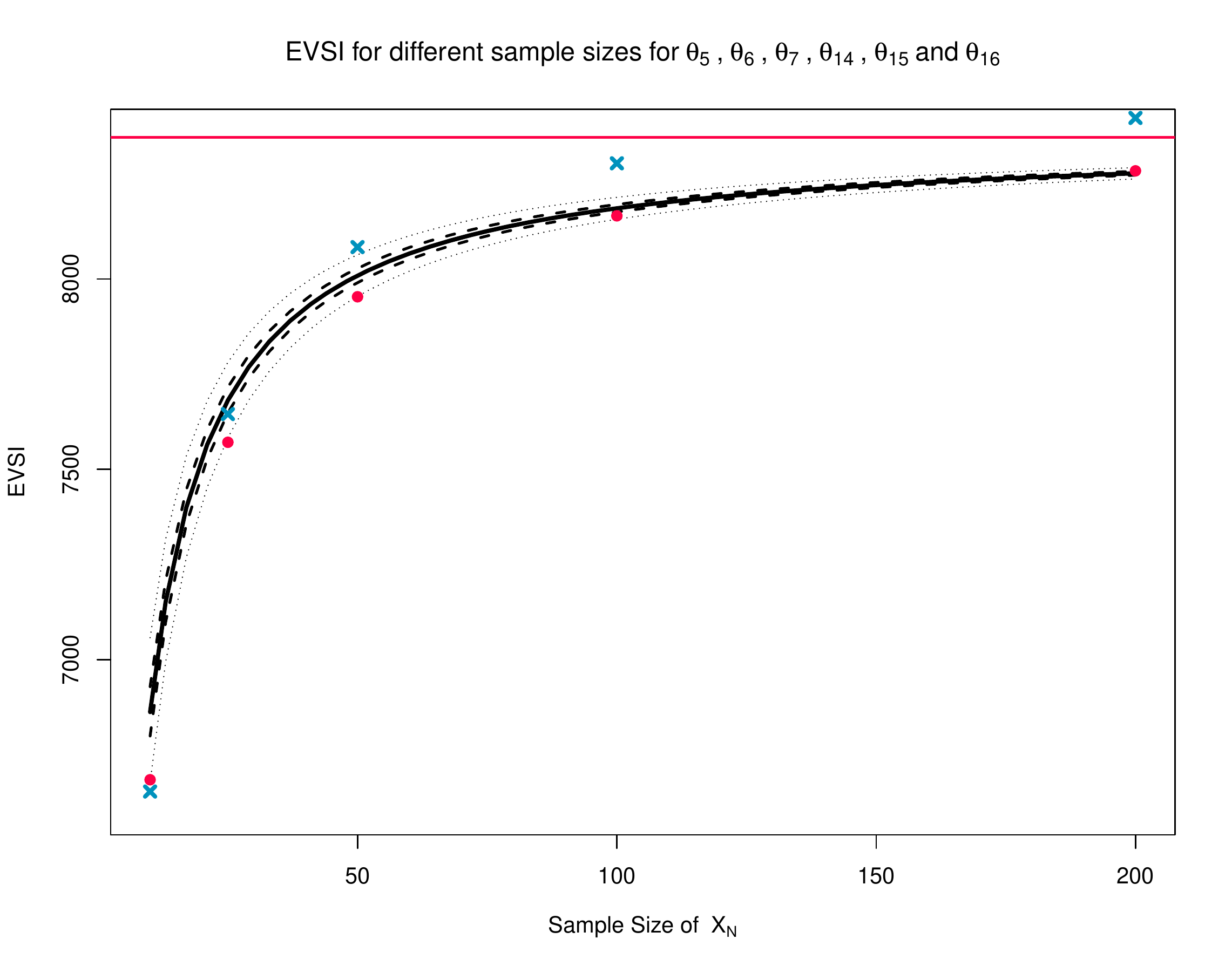}
\caption{The EVSI estimate calculated with the posterior median of $\sigma^2_{\bm X}$ (solid line) along with 75\% (dashed line) and 95\% (solid line) posterior credible intervals for the BK example; Study 1 (Top Left), Study 2 (Top Right), Study 3 (Middle Left), Study 4 (Middle Right) and Study 5 (Bottom). These are compared with the nested Monte Carlo estimates (red dots) and the Menzies estimates (blue crosses). The EVPPI for the parameters targeted by the study is shown as the horizontal red line.}
\label{BK-figure}
\end{center}
\end{figure}

Figure \ref{BK-figure} shows the EVSI estimates for the BK example. The solid line gives the EVSI calculated with the median of the posterior distribution of $\sigma^2_{\bm X}(N)$, whereas the dashed line is the 75\% credible interval and the dotted line the 95\% credible interval. The EVSI estimates from the nested Monte Carlo estimator and the Menzies estimator are given by the red dots and the blue crosses respectively. The nested Monte Carlo estimator (representative of the ``true'' EVSI) is within the 95\% credible interval for all exercises except exercise 5 (bottom), where the EVSI is slightly over estimated for small values of $N$. This small over-estimation may be due to the inaccuracies introduced by estimating INB$^{\bm\phi}$ using only 20\,000 observations, as opposed to the full PSA simulation used for the other examples.

Figure \ref{BK-figure} demonstrates that the EVSI is estimated with more relative precision as the EVSI estimate increases. This is because, for small values of the EVSI, the difference between $\sigma_q^2$ and $\sigma^2$ is small and the estimate of the two variances needs to be very accurate in order to estimate the difference. Therefore, this method for the EVSI calculation should be reserved for situations where the underlying parameters have significant value. If the EVSI estimate is too variable to aid decision making, as seen by the confidence bands, then more simulation should be undertaken. In general, extra simulations should be gained by increasing $Q$, provided the number of posterior simulations is sufficient to characterise the distribution of the ``posterior'' incremental net benefit \cite{Heathetal:2018}.

\vfill

\section{Non-Linear Regression Code}
The following function calculates the EVSI across sample size using the non-linear regression method presented in this paper. It also allows users to plot the confidence bands for the EVSI as suggested in the paper when \texttt{plot=TRUE}.

\begin{verbatim}
calc.evsi.N<-function(INB.phi,INB.theta,sigma.q,N.q,plot=TRUE){
  #INB.phi - The expectation of the INB conditional on phi, normally calculated 
  #          using non-parametric regression
  #INB.theta - The INB conditional on all model parameters
  #sigma.q - A vector of "posterior" variances from simulations
  #N.q - The N.q vector that contains the sample sizes for which the posterior
  #      variance is calculated
  #plot - Logical. Should the graphic be plotted.
  
  model.ab<-c(
    "model{
    #Prior for the non-linear regression parameter
    h~dnorm(h.mean/2,h.tau)  T(0.00000E+00, )
    
    #Non-Linear Model
    for(i in 1:Q){
    sigma.X[i]~dnorm(mu[i],tau)
    mu[i]<-sigma.phi*(N[i]/(N[i]+h))
    }
    #Prior for the residual variance
    sigma ~ dt(sigma.mu, sigma.tau, 3)  T(0.00000E+00, )
    tau <- 1/sigma^2
}")
  
  library(rjags)
  #Writing Model Code into file
  filein<-"model.txt"
  writeLines(model.ab,filein)
  
  #Calculate the variance for the different INB functions
  sigma.INB<-var(INB.theta)
  sigma.phi<-var(INB.phi)
  Q<-length(N.q)
  
  #Set up the data for the jags model
  data.a.b<-list(
    #Parameters for prior of residual variance
    sigma.mu=sd(sigma.INB-sigma.q)/2,
    sigma.tau=1/sd(sigma.INB-sigma.q),
    #Parameters for prior for h
    h.mean=max(N.q),
    h.tau=0.0005/max(N.q),
    #Number of observed variances
    Q=Q,
    #Maximum variance of sigma.x^2
    sigma.phi=sigma.phi,
    #Data for the non-linear model      
    sigma.X=sigma.INB-sigma.q,
    N=N.q)
  
  #Run the jags model with the data
  Model.JAGS<- rjags::jags.model(filein,data=data.a.b)
  update(Model.JAGS,1000)
  beta.ab <- rjags::coda.samples(Model.JAGS, c("h"), n.iter=3000,n.thin=1)
  
  #Calculate the EVSI
  N.fit<-round(seq(min(N.q),max(N.q),length.out=Q))
  quantiles<-quantile(as.matrix(beta.ab),prob=c(0.975,0.75,0.5,0.25,0.025))
  
  #Use these to calculate the quantiles of the fitted sigma.X variances
  quantiles.N<-matrix(NA,nrow=5,ncol=Q)
  for(l in 1:5){
    quantiles.N[l,]<-sigma.phi*(N.fit/(N.fit+quantiles[l]))
  }
  
  #Calculate the EVSI by rescaling INB.phi
  EVSI<-matrix(NA,nrow=5,ncol=Q)
  mu.phi<-mean(INB.phi)
  for(l in 1:5){
    for(k in 1:Q){
      sigma.X<-(INB.phi-mu.phi)/sqrt(sigma.phi)*sqrt(quantiles.N[l,k])+mu.phi
      EVSI[l,k]<-mean(pmax(sigma.X,0))-max(mean(sigma.X),0)
    }
  }
  
  if(plot==TRUE){
    #Calculate EVPPI
    EVPPI<-mean(pmax(INB.phi,0)-max(mean(INB.phi),0))
    #Set up different lines to plot
    lwd<-c(1,2,3,2,1);lty<-c(3,2,1,2,3)
    #Plotting
    plot(as.matrix(N.fit),EVSI[1,],ylim=c(min(EVSI),max(EVSI,EVPPI)),
         type="l",col="white",
         xlab=expression("Sample Size of "~X[N]),
         ylab="Economic value",oma=c(0,0,-1,0))
    for(i in 1:5){
      points(as.matrix(N.fit),EVSI[i,],type="l",
             lwd=lwd[i],lty=lty[i])
    }
    #Add EVPPI to graphic
    abline(h=EVPPI,col="red",lwd=2)
  }
  return(EVSI)
  }
}
\end{verbatim}

\section{Code for the Ades et al. Example}
To give an example of the variance calculation, \texttt{R} code is given below to calculate the $\sigma^2_q$ vector for Exercise 1 in the Ades et al. example. The final line uses the function above to calculate the EVSI using the estimated $\sigma^2_q$ values. The total computation time for this code is around 7 minutes.

\begin{verbatim}
##Ades et al. Model
library(R2OpenBUGS)
library(R2jags)
library(mgcv)
library(MASS)
Adesetal.model<-function(){
  for(i in 1:N){
    X[,i]~dmnorm(theta.5.cor[1:2],Tau.5.1[1:2,1:2])
  }
  
  theta.1~dnorm(10000,1/10^2)
  theta.11~dnorm(15000,1/10^2)
  theta.2~dnorm(0.1,1/.02^2)
  theta.12~dnorm(.08,1/.02^2)
  theta.3~dnorm(5.2,1/1^2)
  theta.13~dnorm(6.1,1/1^2)
  theta.4~dnorm(4000,1/2000^2)
  theta.8~dnorm(.25,1/.1^2)
  theta.17~dnorm(.20,1/.05^2)
  theta.9~dnorm(-.1,1/0.02^2)
  theta.18~dnorm(-.1,1/0.02^2)
  theta.10~dnorm(.5,1/0.2^2)
  theta.19~dnorm(.5,1/0.2^2)
  
  #To give correlated parameters, use multivariate normal dist
  #theta.5,theta.14,theta.7,theta.16
  theta.5.cor[1:4]~dmnorm(mu[],Tau.5[,])
  #theta.6,theta.15
  theta.6.cor[1:2]~dmnorm(mu.6[],Tau.6[,])
  
  #Health Economic Model - effects and costs
  e[1]<-(theta.5.cor[1]*theta.6.cor[1]*theta.5.cor[3])+(theta.8*theta.9*theta.10)
  e[2]<-(theta.5.cor[2]*theta.6.cor[2]*theta.5.cor[4])+(theta.17*theta.18*theta.19)
  
  c[2]<-(theta.11+theta.12*theta.13*theta.4)
  c[1]<-(theta.1+theta.2*theta.3*theta.4)
  
  #Net Benefit with willingness to pay of 100000
  NB[1]<-100000*e[1]-c[1]
  NB[2]<-100000*e[2]-c[2]
}

#Write Model File
filein.model <- "Adesetal.txt"
write.model(Adesetal.model,filein.model)

#Precision Matrices
sig.5<-0.1
sig.14<-0.1
sig.7<-0.5
sig.16<-1
rho<-0.6

S.5<-matrix(c(sig.5^2,rho*sig.14*sig.5,rho*sig.5*sig.7,rho*sig.5*sig.16,
              rho*sig.5*sig.14,sig.14^2,rho*sig.14*sig.7,rho*sig.14*sig.16,
              rho*sig.5*sig.7,sig.14*rho*sig.7,sig.7^2,rho*sig.16*sig.7,
              rho*sig.5*sig.16,sig.14*rho*sig.16,rho*sig.7*sig.16,sig.16^2),
            nrow=4)
sig.6<-0.1
sig.15<-0.05

S.6<-matrix(c(sig.6^2,rho*sig.6*sig.15,
              rho*sig.6*sig.15,sig.15^2),
            nrow=2)
Tau.5<-solve(S.5)
Tau.6<-solve(S.6)

###Data Precision Matrices
sig.5.X<-0.2
sig.14.X<-0.2
sig.7.X<-1
sig.16.X<-2

S.5.X<-matrix(c(sig.5.X^2,rho*sig.5.X*sig.14.X,rho*sig.5.X*sig.7.X,rho*sig.5.X*sig.16.X,
                rho*sig.5.X*sig.14.X,sig.14.X^2,rho*sig.14.X*sig.7.X,rho*sig.14.X*sig.16.X,
                rho*sig.5.X*sig.7.X,rho*sig.7.X*sig.14.X,sig.7.X^2,rho*sig.7.X*sig.16.X,
                sig.5.X*sig.16.X*rho,rho*sig.16.X*sig.14.X,rho*sig.16.X*sig.7.X,sig.16.X^2),
              nrow=4)
Tau.5.1<-solve(S.5.X[1:2,1:2])

#Data to simulate without any data
data<-list(mu=c(.7,.8,3,3),
           Tau.5=Tau.5,
           Tau.5.1=Tau.5.1,
           mu.6=c(.3,.3),
           Tau.6=Tau.6,
           X=NA,
           N=0)

#Global MCMC parameters
n.chains<-3        #Number of chains for MCMC
n.burnin <- 10000  # Number of burn in iterations
n.thin<-20         #Thinning
#1 million simulations without the data
n.iter <- ceiling(1000000*n.thin/n.chains) + n.burnin 

# Choose the parameters in the model to monitor
parameters.to.save <- c("theta.5.cor","NB")

#Use jags to sample from the parameters and find INB.theta
model.theta <- jags(
  data =  data,
  parameters.to.save = parameters.to.save,
  model.file = filein.model, 
  n.chains = n.chains, 
  n.iter = n.iter, 
  n.thin = n.thin, 
  n.burnin = n.burnin,DIC=F) 

#Calculate INB.theta
INB.theta<-model.theta$BUGSoutput$sims.list$NB[,2]-model.theta$BUGSoutput$sims.list$NB[,1]
var(INB.theta)
#Generate N.q and phi.q
Q<-50
N.min<-5
N.max<-200
#Spaced in the sqrt.
N.q<-trunc(seq(sqrt(N.min),sqrt(N.max),length.out=Q)^2)

#phi.q is two dimensional
theta.5<-model.theta$BUGSoutput$sims.list$theta.5.cor[,1]
theta.5.q<-quantile(theta.5,probs=(1:Q/(Q+1)))
theta.14<-model.theta$BUGSoutput$sims.list$theta.5.cor[,2]
theta.14.q<-quantile(theta.14,probs=(1:Q/(Q+1)))

#Reorder phi.q randomly
while(abs(cor(N.q,theta.5.q))>0.001){theta.5.q<-sample(theta.5.q,50,replace=F)}
while(abs(cor(N.q,theta.14.q))>0.001){theta.14.q<-sample(theta.14.q,50,replace=F)}

#PHI FOR THIS EXAMPLE
phi.q<-cbind(theta.5.q,theta.14.q)

sigma.q<-array()
for(i in 1:Q){
  #For this loop set the sample size and phi value for generating the data.
  N<-N.q[i]  
  mu<-phi.q[i,]
  #Generate X conditional on the specific N.q and phi.q
  X<-mvrnorm(N,mu,S.5.X[1:2,1:2]) 
  #For use as data X must be a column matrix
  X<-t(as.matrix(X))
  
  #Use X as data in the model
  data<-list(mu=c(.7,.8,3,3),
             Tau.5=Tau.5,
             Tau.5.1=Tau.5.1,
             mu.6=c(.3,.3),
             Tau.6=Tau.6,
             X=X,
             N=N)
  
  #Moniter the Net Benefit to calculate its variance
  parameters.to.save <- c("NB")
  
  #Use fewer simulations to estimate the posterior variance. 
  n.iter <- ceiling(10000*n.thin/n.chains) + n.burnin 
  
  #Use jags to perform MCMC
  model.posterior <- jags(
    data =  data,
    parameters.to.save = parameters.to.save,
    model.file = filein.model, 
    n.chains = n.chains, 
    n.iter = n.iter, 
    n.thin = n.thin, 
    n.burnin = n.burnin,DIC=F) 
  #Find posterior distribution for the INB
  INB.X<-model.posterior$BUGSoutput$sims.list$NB[,2]-model.posterior$BUGSoutput$sims.list$NB[,1]
  #Calculate the posterior variance
  sigma.q[i]<-var(INB.X)
}

#Find INB.phi
INB.phi<-gam(INB.theta~te(theta.5,theta.14))$fitted.values

####Calculate the EVSI####
EVSI_theta5_theta14<-calc.evsi.N(INB.phi,INB.theta,sigma.q,N.q)
\end{verbatim}

\end{document}